\documentclass[aps,pre,twocolumn,showpacs,showkeys,groupedaddress]{revtex4-2}
\usepackage{graphicx}  
\usepackage{dcolumn}   
\usepackage{bm}        
\usepackage{amssymb}   
\usepackage{subfigure}
\usepackage{amsmath,amsfonts}
\usepackage{setspace}
\usepackage[mathscr]{euscript}
\usepackage[utf8]{inputenc}

\usepackage{xcolor}
\usepackage{hyperref}
\begin{document}
\title{ Entropic bottlenecks  to nematic ordering  in an $RP^{2}$  apolar 
spin model  } 

\author{B. Kamala Latha$^{1}$}
\author{V.S.S. Sastry$^{2}$}
\author{S. R. Shenoy$^{3}$}
\affiliation {$^{1}$School of Physics,University of Hyderabad, Hyderabad 500046, India}

\affiliation{$^{2}$Centre for Modelling, Simulation and Design, University of Hyderabad, Hyderabad 500046, India}

\affiliation {$^{3}$Tata Institute of Fundamental Research, Hyderabad 500046, India}

\date{\today}%

\begin{abstract}

\noindent The Lebwohl-Lasher model of uniaxial liquid crystals with 
(\textit{n} = 3, \textit{d} = 2)  was reported earlier to undergo a crossover 
transition to a novel nematic phase at a temperature  $T=T_{n}$. This phase has unbound 
topological defects in a nematic background, that pair at a lower 
$T_{\text{BKT}} < T_{n}$. The transition has zero latent heat, and a 
specific heat and correlation length that remain finite. We discover here 
a significant sparseness of states or an entropy barrier `bottleneck', between 
the isotropic and novel nematic phases. Passage through these sparse 
configurations is enabled by short-range nematic clusters dressing the 
defect cores. The free energy temperature derivatives, along with energy 
derivatives of the micro-canonical entropy, determine that  this is a 
{\it third-order} transition in the Ehrenfest classification. The local 
transformation to dressed defects induces a sharp downward cusp in the 
correlation length, at a precursor temperature $T_{p} > T_{n}$. The  
entropic bottleneck manifests as a rippling of the  free energy landscape, 
over mutually modifying nematic order and defect density. Cooling 
through  $T_{p}$ yields an itinerant para-nematic fluid of dressed 
defects with macroscopically occupied local polar angle tilts, that 
catalyse a common global tilt or nematic phase at $T_{n}$.
   \end{abstract}

\maketitle

\section{Introduction}
Sustained interest in two dimensional ($d=2$)  lattice models 
with \textit{n}-dimensional apolar spin manifolds at each site ($RP^{n-1}$
systems), arises from the delicate interplay between stable topological
point defects and spin ordering mechanism \cite{Solomon,Kunz,Catterall,
Chiccoli,Mondal, Dutta04,  Berche, Paredes,Farinas-Sanchez, Tomita, 
Shabnam,BKLPRL, Ozeki, Delfino, Bonnati,Diouane, Ueda,Burgelman}. 
The $RP^{2}$  model (\textit{n} = 3, \textit{d} = 2),  which has a  
non-trivial first fundamental group
of its order parameter (OP) space  $\Pi_{1} = Z_{2}$  \cite{Mermin},
hosts stable $\pm 1/2$ topological defects. Unlike its \textit{n} = 2 counterpart 
(formally equivalent to the 2DXY model with a Berezinki-Kosterlitz-Thouless (BKT)
 type transition  \cite {Berez2,KT,Kosterlitz,Young}), the nature of the transition
in $RP^{2}$ model is not settled. Early Boltzmann Monte Carlo 
 simulations  using Metropolis  algorithm \cite{Metropolis} of this
system  found a   BKT-type transition \cite{Kunz,Chiccoli,Mondal,
  Dutta04, Berche,  Shabnam}. However accumulated evidence points 
  to a variety of other conclusions, including a first-order 
  transition, a crossover to a zero-temperature transition, a new universality 
  class without a critical line, a non-divergent spin (apolar) correlation 
  length, or a crossover to a   novel nematic phase \cite{BKLPRL, Ozeki, 
  Delfino, Bonnati,Diouane, Ueda,Burgelman}.
   
   In this paper we examine the formation and nature of the ($n=3$) 
   novel nematic phase \cite{BKLPRL}, continuing with the Monte Carlo 
   sampling procedure  supplemented by  the density of states (DoS), 
   computed with the Wang-Landau  algorithm \cite{WL,LW,Lbook}. 
   Henceforth, we  refer to this method as the entropy augmented 
   Monte Carlo protocol (EAMC) to distinguish it from the conventional 
   Metropolis based Boltzmann Monte  Carlo protocol (BMC). The EAMC 
   protocol  finds formation for $T \leq T_{n}$ of a novel nematic phase. 
   In this case the ordered phase is a global nematic background 
   interpenetrated by unbound topological defects  (that  bind in pairs 
   only at a lower $T_{\text{BKT}}$ Whereas the BMC 
   protocol only yields unbound defect binding in pairs at a 
    $T_{\text{BKT}}$, without any characteristic preceding 
   temperature \cite{BKLPRL}.

Studying the equilibrium emergence of the novel nematic, we find here 
that the EAMC protocol for $(n =3$) reveals a surprisingly rarity of transitional 
microstates in the configuration space between ordered and disordered 
regions. The EAMC enables passage through these configurational 
bottlenecks or entropy barriers, and introduces another characteristic 
crossover temperature $T=T_{p} > T_{n}$, where bottleneck passage 
commences. The two temperatures $T_{p}, T_{n}$ induce complementary 
behaviour in physical quantities. The (apolar)  spin correlation length, 
sharply deviates downwards from its divergent high temperature path, 
inducing a strong upward cusp at $T_{p}$, but it only shows a soft 
slope change at $T_{n}$.  Whereas the specific heat has soft curvature 
maximum that marks $T_{p}$, but has a prominent upward pointing cusp 
at the lower transition temperature $T_{n}$.
We find from computing the canonical free energy as a function of 
temperature, that  the third derivative has a jump at $T=T_{n}$: there 
is manifestly, a third order transition in the Ehrenfest classification. 
Supplementary evidence, from analysing different orders of  energy derivatives 
of the micro-canonical entropy, shows the transition does not have 
signatures \cite{Gross,Schnabel,Qi} of first- or second-order 
transitions. To understand the configurations that facilitate entropy 
barrier passage, we study post-quench transformation transients 
through an EAMC protocol adapted to follow free-running MC evolutions, 
as occur in the partial equilibration scenario (PES) \cite{Ritort,Bonilla,Garriga,Crisanti04,Crisanti13,ShankarEPL}. 
Seemingly unimportant short-range correlations between nematic clusters 
and defects can become surprisingly relevant, in crossing the sparsity 
gap between initially isotropic and finally nematic configurations.

The role of the extra spin dimension of this (\textit{n} = 3) model \cite{BKLPRL} 
is clarified by a spherical harmonic expansion of the Hamiltonian, showing 
qualitatively different contributions by azimuthal and polar angles. The 
azimuthal angle appears only as a nearest neighbour difference variable 
across 2D lattice bonds, while the polar angle appears only as a local 
variable on 2D lattice sites. Just as for ($\textit{n}=2$) case, the 
azimuthal difference fluctuations will suppress azimuthal long range 
order and nucleate azimuthal defects in pairs \cite{MWH1,MWH2,MWH3,Rice,
Straley,Frenkel,Barma}. We find from EAMC simulations yielding distributions 
of polar and azimuthal angles, that canonical averages of director 
orientations change on cooling from  polar angles fluctuating randomly 
about the equatorial plane, to condensing around special out-of-plane 
polar values. The polar and azimuthal spectra narrow on cooling to 
sharper macroscopic occupations, in an analogue of Bose 
condensation, commenced at $T_{p}$ and completed near $T_{n}$.

 The plan of the paper is as follows. Section II  describes the model,  
 briefly presents details of simulation, and   introduces relevant observables. 
Results are presented and  discussed in the subsections of Section III,
 while Section IV  gives a summary and mentions further work. 
 
\section{Hamiltonian and Details of Simulation}
 Physical realizations of $RP^{2}$ symmetries include  2D fully-frustrated 
 Heisenberg antiferromagnet on a triangular lattice 
  \cite{Kawamura1,Kawamura2,Everts}, and  thin film of   uniaxial liquid crystal. 
 The  latter is  described by the two-dimensional Lebwohl-Lasher 
 Hamiltonian (or 2D LL model)  \cite{Saupe,LL,Zannoni},
 \begin{equation} 
 H   = - \epsilon \sum_{i j}  P_{2} (\cos \gamma_{ij}).
 \label{eqn:1}
 \end{equation}
Here  $P_{2}$ is the second Legendre polynomial, $\gamma_{ij}$ is 
the angle between the 'spins' at \textit{i} and \textit{j} sites on a  lattice
with \textit{n} = 3.  Here the energy $\epsilon $ scales $T$ in reduced units. 

The EAMC protocol based on the Wang-Landau algorithm \cite{WL,LW},
augmented by the \textit{frontier sampling} technique \cite{Zhou, BKL15},
was used in the earlier study \cite{BKLPRL}. The details of the procedure as 
adopted to liquid crystals are given in \cite{Jayasri05, BKL15,BKL18}.
The EAMC simulations are done on  a square lattice of size $N= L^{2}$ (\textit{L} = 128), with periodic
boundary conditions. The  protocol  determines the bin DoS $g(E_{\mu})$  of a  
 contiguous set of very thin  system energy bins $\{\mu\}$, and hence bin 
 configurational  entropy $S_{\mu} (E_{\mu}) =  \ln g(E_{\mu})$ which is 
 identified with the micro-canonical entropy of the system. 
The inter-bin  entropy-slope is 
$\beta (E_{\mu}) =   dS_{\mu} /dE_{\mu} = 1/ T_{\text{eff},\mu}$. 
 Noting that $T_{\text{eff},\mu} = T_\text{eff} (E_\mu)$, we suppress
the bin subscript for $ T_\text{eff}$ and 
$ \beta_\text{eff}$ in the following 
discussion, unless explicitly required. We identify $T_\text{eff}$ as 
micro-canonical temperature associated with the  bins.

An entropic ensemble of $ \sim 10^{8}$ microstates (uniformly distributed 
over energy) is generated by effecting the system to perform a random walk in  
configuration space, guided by acceptance probability determined by the
 inverse of the computed DoS. Equilibrium 
ensembles (comprising of $\geq 10^{6}$ microstates) at the chosen 
temperatures are extracted from this collection of microstates using 
a reweighting procedure \cite{Swendsen, BKL15}, termed as RW 
ensembles \cite{BKLPRL}. Computed  observables include system  energy $E$ ($e = E/N$, per site),
  system  entropy S ($s = S/N$, per site), specific heat per 
 site $C_{v}$, nematic order parameter $S_{n}$ and its susceptibility 
 $\chi$, unbound defect density $\rho_{d}$ and  topological parameter $\delta$.
 In the following,we examine the variations of the observables both
 in equilibrium as well as at the micro-canonical (bin) level.  
   Specific heat is related to energy fluctuations as 
    $C_{v} = (<E_{c}^{2}> - <E_{c}>^{2}) / N T^{2}$
where $E_{c}$ is the  configurational energy of the microstate.  The 
equilibrium value of the nematic order parameter $S_{n} = \langle S_{n, c} \rangle $ 
is the ensemble  average of the orientational (apolar) order of constituent 
configurations. Here $S_{n,c} $ is computed
as  the average of projections of site directors of the configuration 
$C$, along the preferred orientation of its nematic  director. The nematic susceptibility is 
  $\chi = (<S_{n, c}^{2}> - <S_{n, c}>^{2})/T$, with computational details in \cite{Zannoni}. 

  To calculate the  topological parameter $\delta$ and 
   unbound  defect density $\rho_{d}$,  the procedure outlined in  
   \cite{Kunz,Dutta04,BKLPRL} is adopted:
 A unit vector  $\bm{\sigma(r)}$ is assigned at each site $\bm{r}$ on the square lattice 
 $\mathcal{L}$  representing  the local site director orientation. Their orientations 
 are represented  by points on  the unit sphere in the order parameter (OP) 
 space $\mathcal{R}$.  Unit sphere with  antipodal points identified 
 correspond to the $RP^{2}$ projective plane. Considering  $\bm{\sigma(r)}$
 and $\bm{\sigma(r^{'})}$ at  two neighbouring lattice sites, $(\bm{r}, \bm{r}^{'}$), 
 we assign a path on the surface of this sphere in  $\mathcal{R}$
 by choosing the shortest geodesic connecting them. Any closed loop 
 on  $\mathcal{L}$ is thus mapped  to a loop $\mathcal {W(L)}$ in the 
 OP space.  The homotopy class of this map is given by 
$\mathcal{W(L)}= \prod_{(\bm{r},\bm{r^{'}}) \in \CMcal{L}}~ 
{sgn}(\bm{\sigma(r)}, \bm{\sigma(r^{'})})$, where the product is 
sequentially ordered over $\mathcal{L}$ and \text{sgn} operates on 
the inner product of the two vectors. 
Topological order $\nu$ of the configuration  is computed as the 
 average of $\mathcal{W(L)}$ on  closed loops generated by the toroids 
over the lattice (making use of periodic boundary conditions). 
The topological parameter  $\delta$  is defined as $\delta= (1-\nu)/2$. 
The density of  unbounded charge 
 $1/2$ defects $\rho_{d}$ for a given configuration is computed by dividing the 
 lattice into a composition of elementary triangular plaquettes.
 The above product applied to each plaquette yields a defect finding 
 algorithm: if the ordered product is -1, the plaquette 
 encloses a charge $1/2$ defect. The  defect density of a configuration
  is calculated from the relative count of such isolated 
 defects over all triangular plaquettes of the  lattice. The equilibrium 
 values of $\delta$ and $\rho_{d}$ are obtained from their 
 ensemble averages \cite{Kunz,Dutta04, BKL20}. 
These data are computed as a function of temperature
  in the  range  [0.05, 1.05] with a temperature resolution of 0.001.
 
 The canonical  free energy  per site $f(T)$ is calculated using 
  Legendre transformation on equilibrium entropy per site  \cite{Callen}. 
  The spatial variations of the  orientational pair correlation function 
  $ G(r) = G(r_{ij}) \equiv \langle P_{2} (\gamma_{ij}) \rangle$
  is computed at 80 temperatures in the above 
  temperature range. Statistical errors, estimated with the  
  jack-knife  algorithm \cite{BKL16}, in $E$, $S_{n}$,
  $ \delta$ and $\rho_{d}$ are typically of the order 
  of 1 in $10^{3}$, while higher moments ($C_{v}$, $\chi$),
   are relatively less accurate (about 5 in $10^{2}$).    

 In this study we improved the accuracy of  convergence of the DoS
by making a modification to the algorithm. Small entropy barriers 
that are introduced as part of this protocol, to discriminate temporarily
 against the already converged high entropy regions (at the so called 
"frontiers'') \cite{BKL15}, are now made a function of energy, - 
(kept constant in the earlier work).  This 
resulted in an improvement of  the degree of convergence of the DoS, with
the  tolerance limit reducing  from  $10\%$ to  $3\%$.
 Consequently, there is a uniform shift of temperature to higher side  by 
 $+0.021$ (relative to  \cite{BKLPRL}),  
 with no change in  the intervals between the
   characteristic temperatures within  computational errors.
  
   A  new algorithm was  developed to track the non-equilibrium
   pathways of the system during a  temperature quench simulation. The Markov chain
   sequence of this evolution is constructed by applying to the trial state,
   an acceptance criterion based on both entropy and energy increments
   induced by the random steps. Each equilibration pathway is a realization
   of  a Markov sequence so constructed. The values of relavant observables 
   are now functions of the Monte Carlo time, measured in units of 
   lattice sweeps (Monte Carlo Sweeps, MCS). Adaptation of the quench protocol 
   to equilibrate the sample with the EAMC protocol  facilitated 
   construction of large equilibrium ensembles at different temperatures. 
   The algorithm is  extended to guide the system through a series of 
   equilibrium  ensembles of various sizes, at  closely spaced temperatures,
   starting either from a high- or low-temperature arbitrary initial state. Thus the 
   quasi-static pathways of the system, during cooling or heating cycles,
   are  tracked with the EAMC procedure.  
         
\section{Results and Discussions}

An energy-uniform random walk  over the energy interval bracketing the 
crossover \cite{BKLPRL} is effected by biasing it with the inverse of the DoS. 
Several observables of the sampled configurations are recorded over 
$10^{6}$ sampled microstates. The range of each observable is divided
into 100 uniformly spaced bins. The random walk data are sorted in the 
form of  histograms pertaining to selected variable pairs, or the two different  
microstate observables of interest, say $(A,B)$. In representative plots, such 
a histogram would appear as the continuous curve of a (joint) probability density 
$p$ of the microstate pair $(A,B)$. We will compactly term this function $p(A,B)$ 
as the density of microstates,  presented as 3D mesh plots 
projected over the planes of the microstate variable pairs. Different 
examples  of the density of microstates
 are given in Figs. 1, 3, 4 below. Free energy surfaces can be obtained 
 over the same microstate  variable pairs as shown  in Figs. 8, 9, 10. 
 Curves of system properties versus control variables can also be 
 derived from appropriate projections.

\subsection{Entropy bottlenecks}

\begin{figure}[h!]
\centering 
\includegraphics[width=0.5\textwidth]{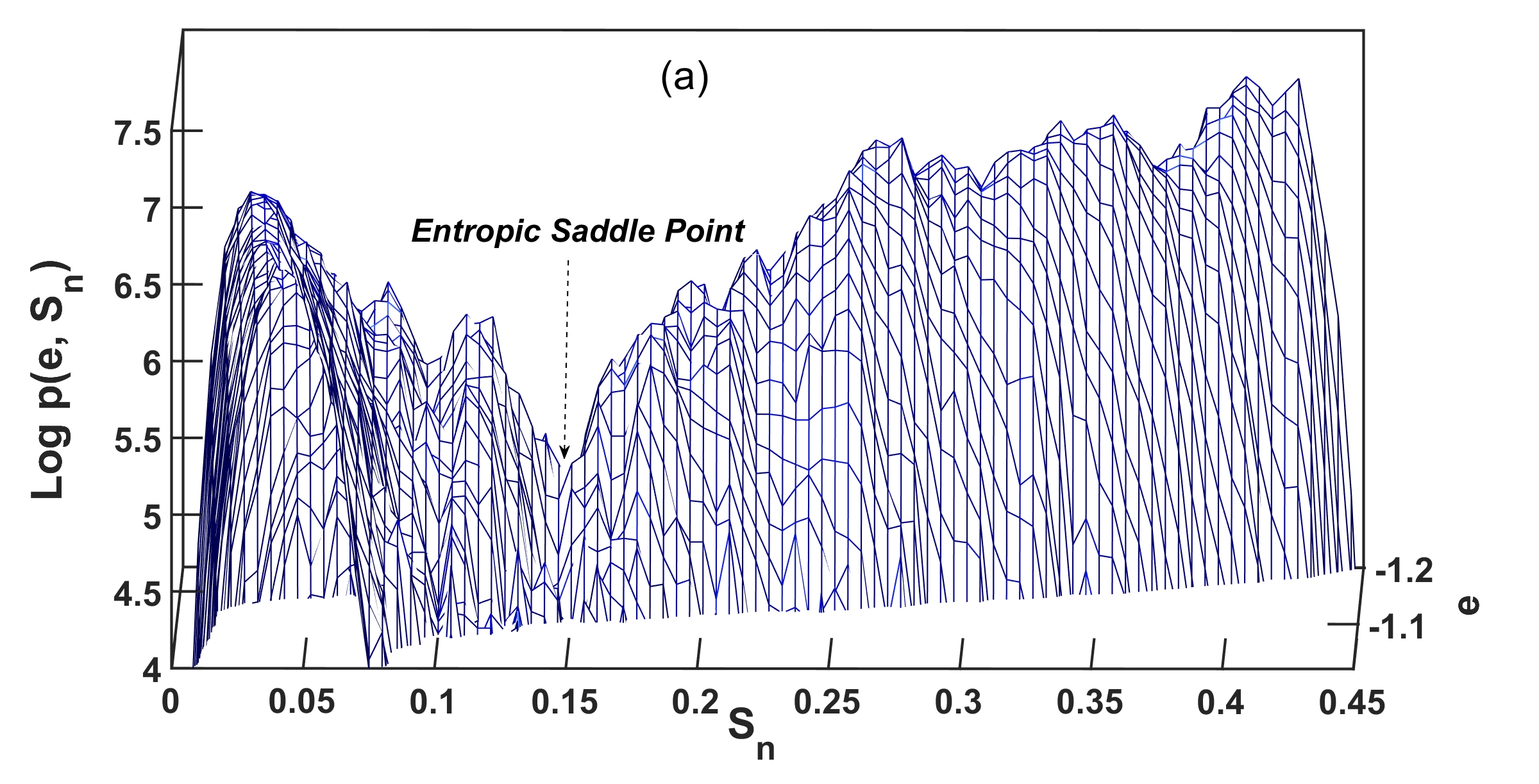}
\includegraphics[width=0.5\textwidth]{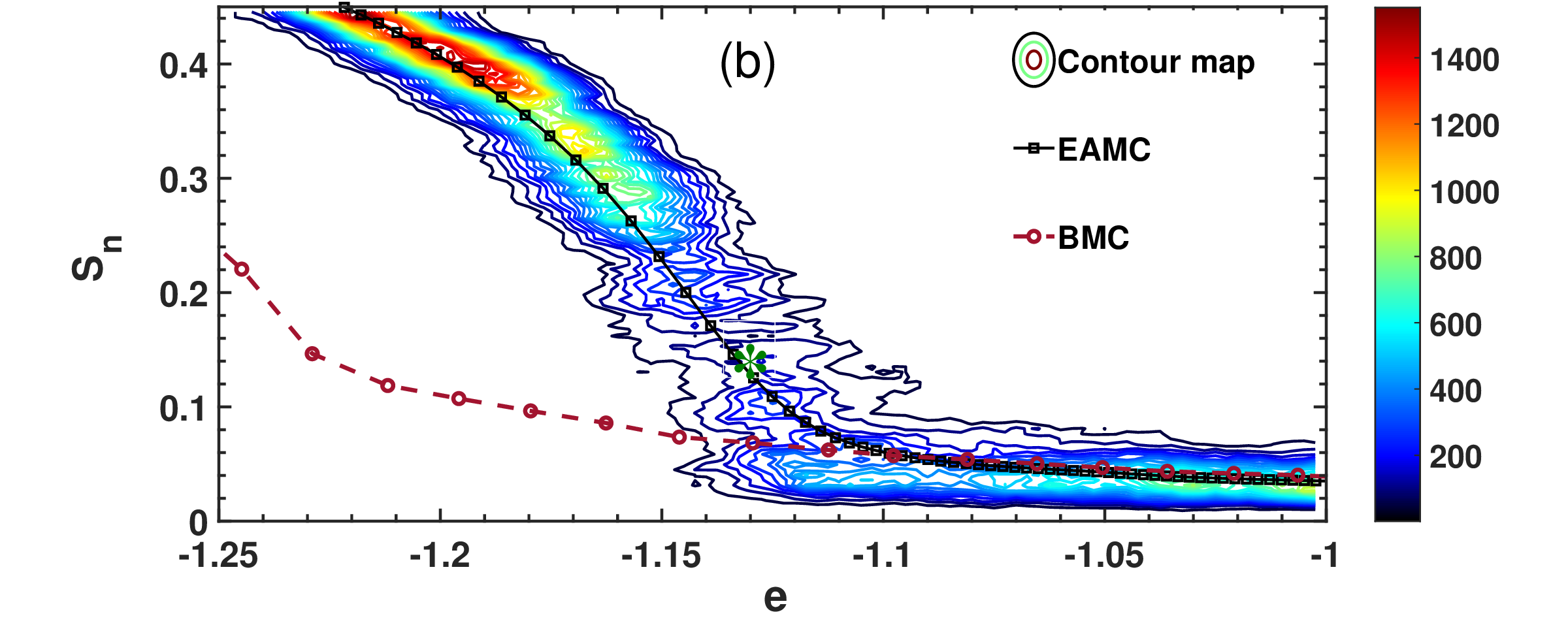}
\caption{(color online) {\it Sparseness of states between two special energies: }
(a)  Data recorded during an energy-uniform random walk, shown as a 
3D mesh plot of the logarithm of the density of microstates
$\log p(e,S_n)$, over the plane of energy and order parameter 
$(e,S_n)$. There is  an entropic saddle-point near $(e, S_n)= (-1.122,0.14)$. 
Over the interval from tunnel entry at $(-1.1, 0.05)$, the entropy drops 
by an order of magnitude.
(b) The density of microstates $p(e, S_n)$ shown as a 
contour plot  on the plane of  ($e, S_{n}$). Superimposed on the plot 
are canonical-equilibrium averages from BMC (dashed  line) and 
EAMC protocol (solid line) that match in the initial  high energy region,
Below some $e \simeq -1.11$, the BMC pathway parts company 
from the EAMC, that  traverses a sparse microstate region to find 
richer states at a lower $e_{n} \simeq -1.16$.}
\label{fig:1}
\end{figure} 

 Fig.~\ref{fig:1}(a) depicts a 3D mesh plot of the density of microstates
 as a logarithmic  $\log p(e, S_{n})$ over the selected plane of variable 
 pairs of energy per site $e$ and nematic order parameter $S_n$. 
 The minimum height of the entropic tunnel identifies the rarest 
 configuration space, or highest entropy barrier, near $(e,S_{n})  = (-1.122, 0.14)$.

  Fig.~\ref{fig:1}(b)  presents the same $p (e, S_{n})$ data as contours 
  projected on  the ($e, S_n$) plane. There is a significant  
  {\it sparseness} between two special energies, with population-rich 
segments on either side. Superimposed on the contours are
plots of quasi-statically guided canonical averages of  ($e, S_{n}$) 
dashed and solid lines from BMC and EAMC protocols, respectively. 
The lines match at high energies, but part company and evolve 
differently over energies  below a transformation energy, when
the EAMC protocol enables the system to traverse  the sparse regions 
to find a novel nematic phase. 
The BMC protocol is seen to differ from this path, and has been 
 reported to proceed to a BKT-type transition. 
 
The polar angles of the $(n=3)$ rotations seem to contribute to the branching
 characteristics of the model. The non-Abelian nature of the  $SO(3)$ group 
 operators \cite{Kunz}   is plausibly responsible for the relatively small 
 number of transition pathways.  For the $(n=2)$ case, a local isotropic 
 region can be transformed to a nematic cluster by a sequence of 
 azimuthal angle increments of the site directors. The connecting 
 configurations are dense, as an exponentially large number of 
 equivalent pathways can be generated, by shuffling the given 
 sequence of incremental Abelian rotation operations.
For the $(n=3)$ case however, a random shuffling
of  non-Abelian angle increment operations will not in general generate  
the same nematic  cluster: only a particular sequence of operations 
will do this. The pathways to access  this branch with nematic long range order,  are 
hence necessarily {\it sparse}.
    
\begin{figure}[h!]
\centering
\includegraphics[width=0.5\textwidth]{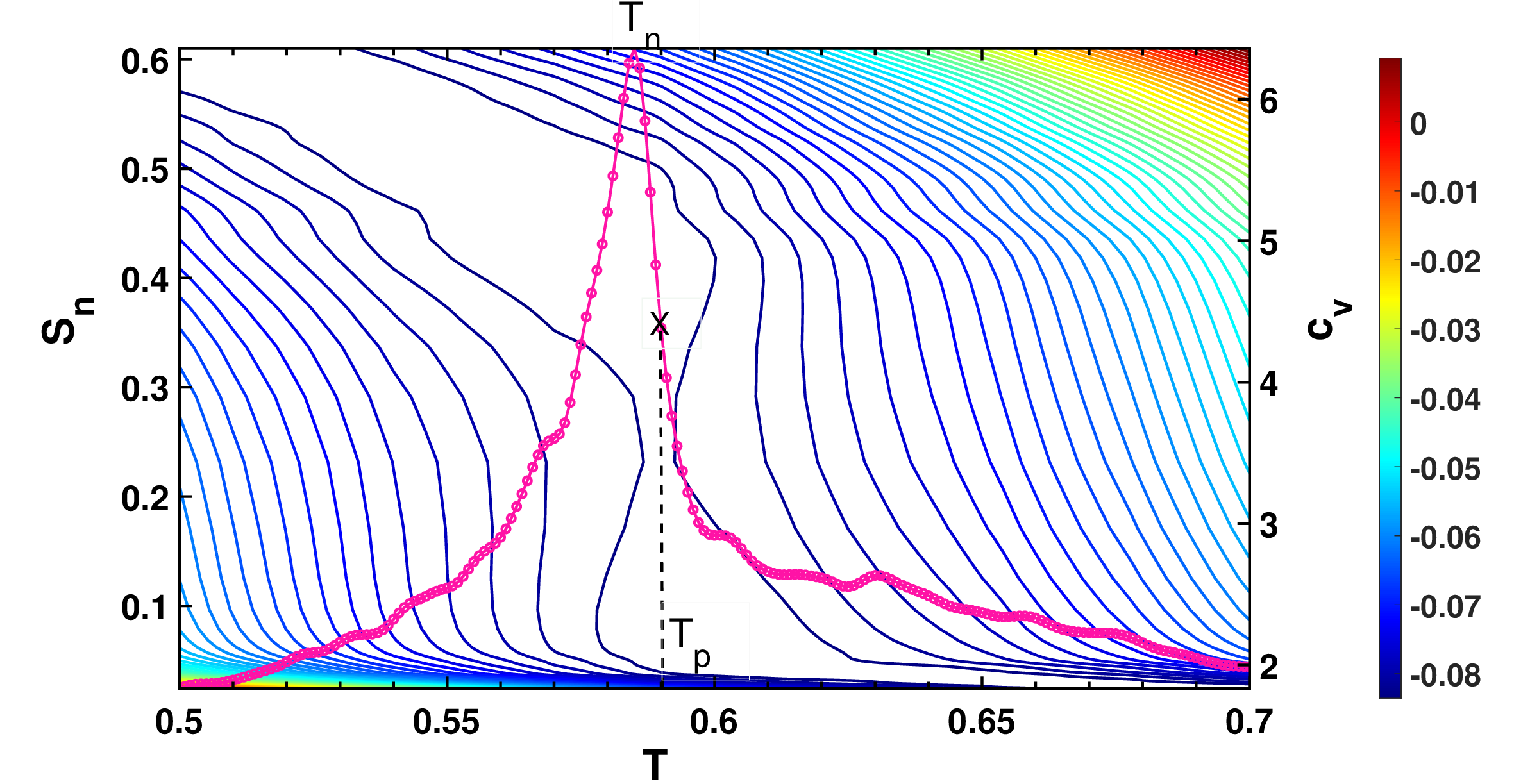}
\caption{(color online){\it  Two special temperatures: } For $L = 128$, the 
3D plot of  free-energy per site $f(S_{n}, T)$ is projected as a contour map on  the 
equilibrium $(S_{n}, T )$ plane, with a temperature  resolution $\Delta T$ = 0.001, showing a 
narrow contour bottleneck. The superimposed $C_{v}(T)$ (red  line with 
symbols) shows a peak at $T = T_{n} = 0.585$. The  maximum-curvature point 
on the high temperature side (vertical dash line through the bottleneck) 
defines a precursor temperature $T= T_{p} = 0.590  (> T_{n})$.}
\label{fig:2}
\end{figure}  
Fig.~\ref{fig:2}  is the contour plot of the  EAMC-derived canonical 
 free energy per site $f(S_{n},T)$, projected on the equilibrium  
 ($ S_{n}, T$) plane. Also  superimposed on this data is   $C_{v} (T)$ per site 
 with a  cusp at  $T = T_{n}$.   There is  a  narrow 
contour-constriction seen  at the bottleneck centered at  
$T = T_{p}$, where there is a maximum curvature $C_v (T)$ point, preceding $T_{n}$.   

\begin{figure}[h!]
\centering
\includegraphics[width=0.5\textwidth]{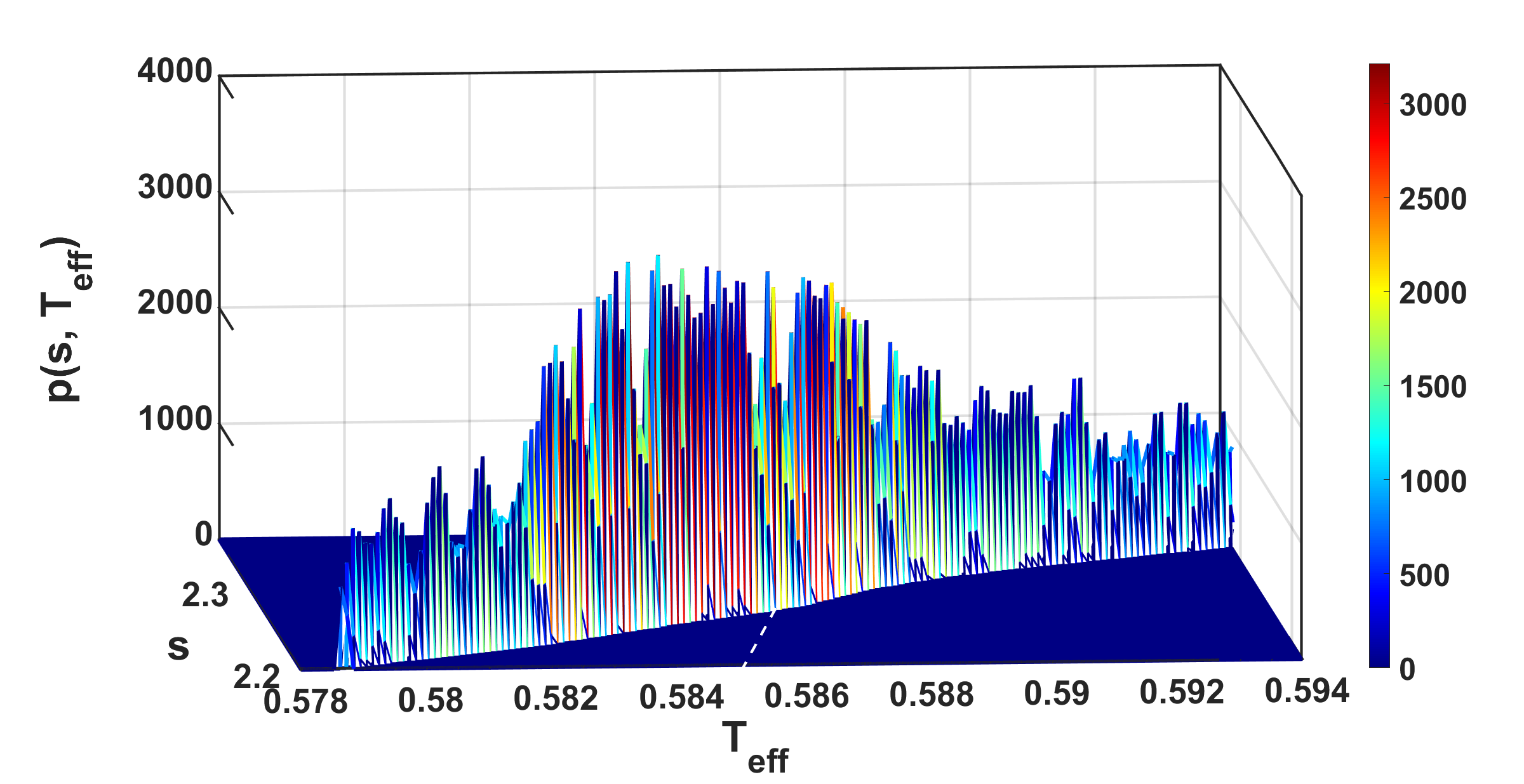}
\caption{(color online) {\it Density of microstates over entropy and temperature: } 
3D mesh plot of  density of microstates
$p (s,T_\text{eff})$  over the plane of the entropic variable pair 
($s, T_\text{eff}$) in the entropy barrier region. There is an 
absence of latent heat (no jump in $s$) at the transition 
temperature $T_{n}= 0.585$ (dashed line). The bottleneck 
region shows an enhanced exploration. }
\label{fig:3}
\end{figure} 
The density of microstates$p(s,T_{\text{eff}})$ 
in Fig.~\ref{fig:3} shows a visible accumulation of visited states 
reflecting a sustained exploration  by the system for connecting 
pathways while exploring the important bottleneck region.
 
\begin{figure}[h!]  
\centering
\includegraphics[width=0.5\textwidth]{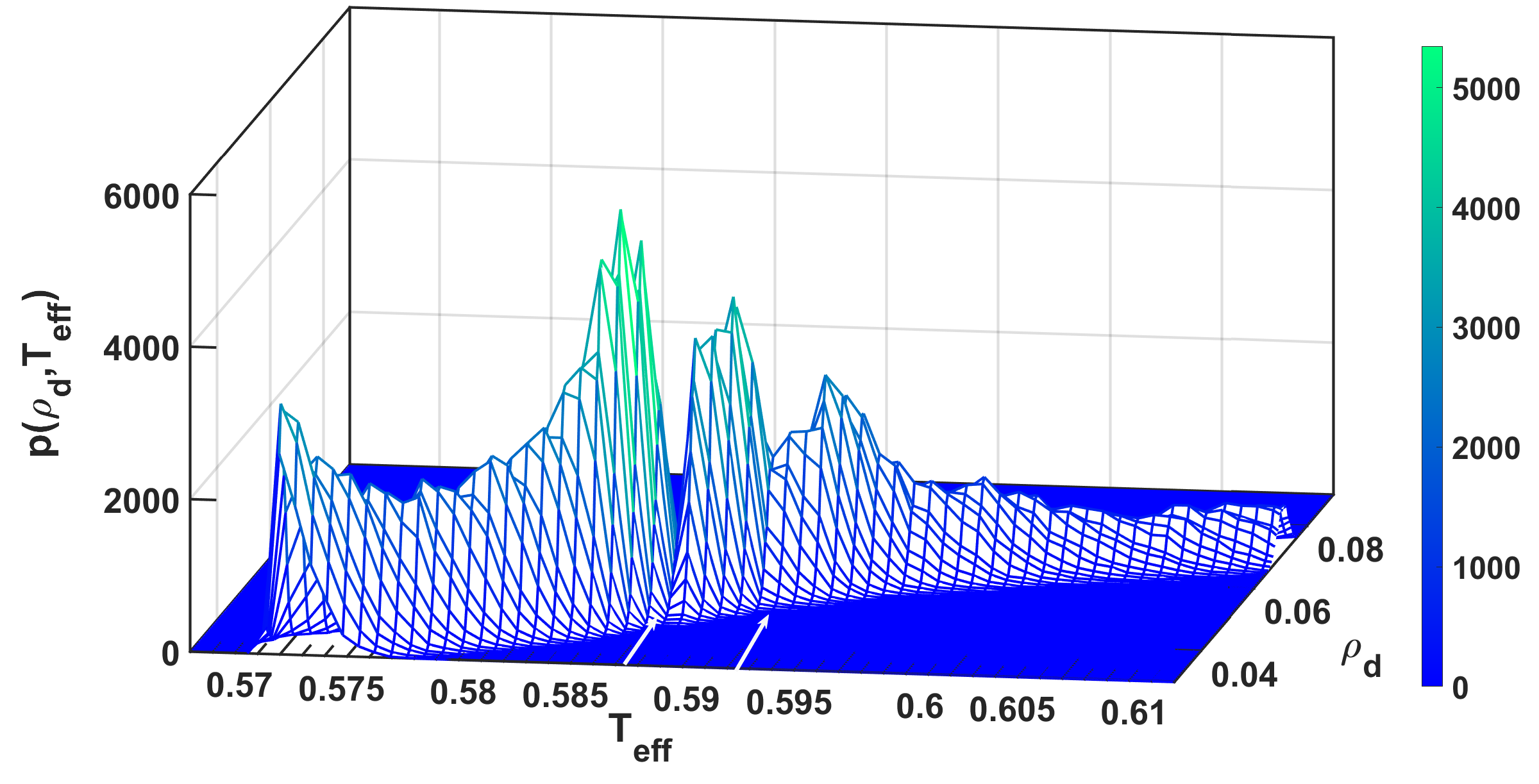}
\caption{(color online) {\it  Density of microstates over defect density and temperature: }
 3D mesh plot of the density of microstates $p(\rho_{d},T_{\text{eff}})$ 
 over the plane of defect density and temperature in the entropy barrier region.
Dips (indicated by arrows) are at the {\it same} $T_{p}, T_{n}$ as previously, 
showing correlations between  defect cores and short-range nematic order.}
\label{fig:4}
\end{figure}  
Fig.~\ref{fig:4} shows the data of Fig.~\ref{fig:1}, now projected  as 
the density of microstates
 $p(\rho_{d}, T_{\text{eff}})$ in the plane of temperature $T_\text{eff}$ and 
 defect density $\rho_{d}$.  It has dips  at the OP-related temperatures 
$T=T_{p}$ and $T_{n}$, showing that temperature variations of $S_{n}$ and $ \rho_{d}$ 
are correlated. This correlation is also  seen as evident  in the equilibrium free energy 
profiles (presented later).

\begin{figure}[h!]
\centering
\includegraphics[width=0.45\textwidth]{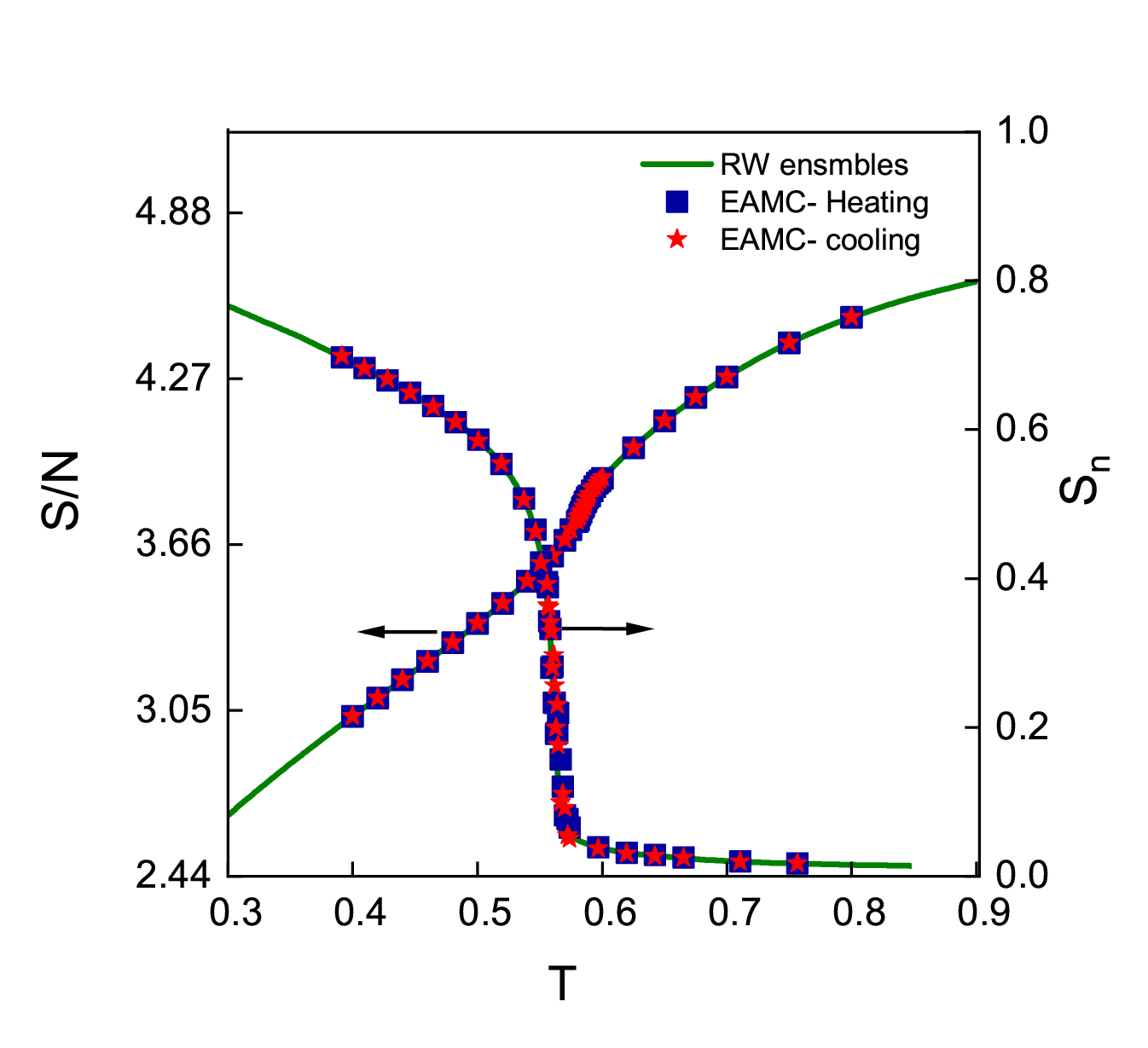}
\caption{(color online) {\it Absence of hysteresis at transition: } Equilibrium 
variation of nematic order and canonical entpy per
site s, bracketing the transition at $T_{n}$.
Data on both, collected during cooling and heating cycles, overlap respectively,
 ruling out hysteresis. The solid lines indicate equilibrium data 
 from reweighting procedure of the random walk ensemble.}  
\label{fig:5}
\end{figure}

In Fig.~\ref{fig:5} we present the equilibrium data on 
order parameter $S_{n}$ and canonical entropy per site $ s$ 
in the temperature range [0.4, 0.8], collected during heating and cooling 
cycles. The data overlap without hysteresis, implying that the 
nematic transition is not first order.

\begin{figure}[h!]
\centering
\includegraphics[width=0.5\textwidth]{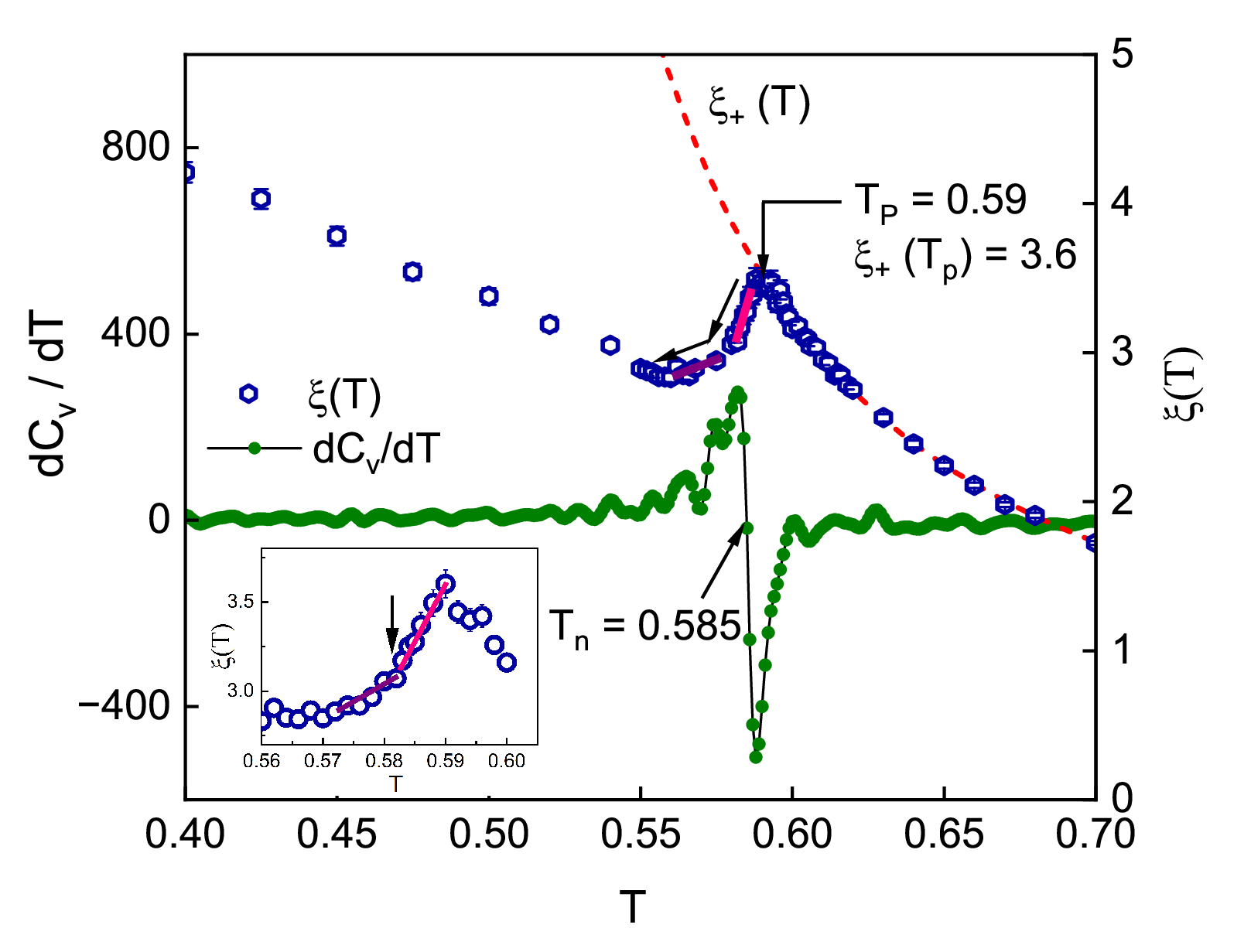}
\caption{(color online) {\it Structure in specific heat and correlations 
at the two special temperatures:} 
 The specific heat  slopes $dC_{v} /dT$ are of opposite sign on 
either side of $T=T_{n}$, arising from  a sharp $C_{v}$ cusp. The short-range 
spin correlation length  $ \xi (T)$, initially locked to the rising 
defect-core separation  $\xi_{+}(T)$ (red dash line), peels off as a 
sharp cusp at $T= T_{p} = 0.590$, when its value is  $\sim 3.6$. The region 
of a slope change indicated by two arrows, is magnified in the Inset, where
 $\xi (T)$  has  a weak slope change at $ T \simeq T_n$ (downward arrow).}
\label{fig:6}
\end{figure}
The orientational spin correlation function $G(r)$  facilitates calculation 
of the characteristic length associated with nematic director fluctuations. 
With the EAMC protocol, we find $\xi (T)$ from fits to 
$G(r ; T) -{S_n (T)}^2   \sim e^{-r/ \xi(T)}$. Fig.~\ref{fig:6} shows that for 
$T \geq T_{p}$, the ($n=3$) short-range OP correlation  length $\xi (T)$ 
in this temperature range, is locked to a BKT-like $\xi_{+} (T)$ diverging 
form  (as in $n = 2$ case) corresponding to an essential singularity \cite{Kosterlitz}. 
Here $ \xi_{+}(T) = A_{0} e^{[A_{1} / (T - T_{\text{BKT}})^{1/2}]}$, 
with fitted constants 
$T_{\text{BKT}} = 0.413$, $A_{0} = 0.129$,  and $A_{1} = 1.392$. At $T = T_{p}$, 
the temperature variation of $\xi (T)$ changes qualitatively. As 
seen in  Fig.~\ref{fig:6}, $\xi (T)$ decreases sharply below $T_{p}$, forming a cusp. 
This qualitative change for $T < T_{p}$ in the nature of $\xi (T)$ is recognised, by 
writing it as the nematic correlation length $\xi_{n}(T)$.
 The inset of Fig.~\ref{fig:6} shows that the  variation of $\xi_{n}(T)$ below 
 $T_p$ has softer slope change near the nematic transition temperature 
 $T \simeq T_{n}$. On further cooling $\xi_{n}(T)$ increases as the nematic order consolidates. 

\subsection{ Acceptance probabilities of different protocols }
 \begin{figure}[h!]
\centering
\includegraphics[width=0.5\textwidth]{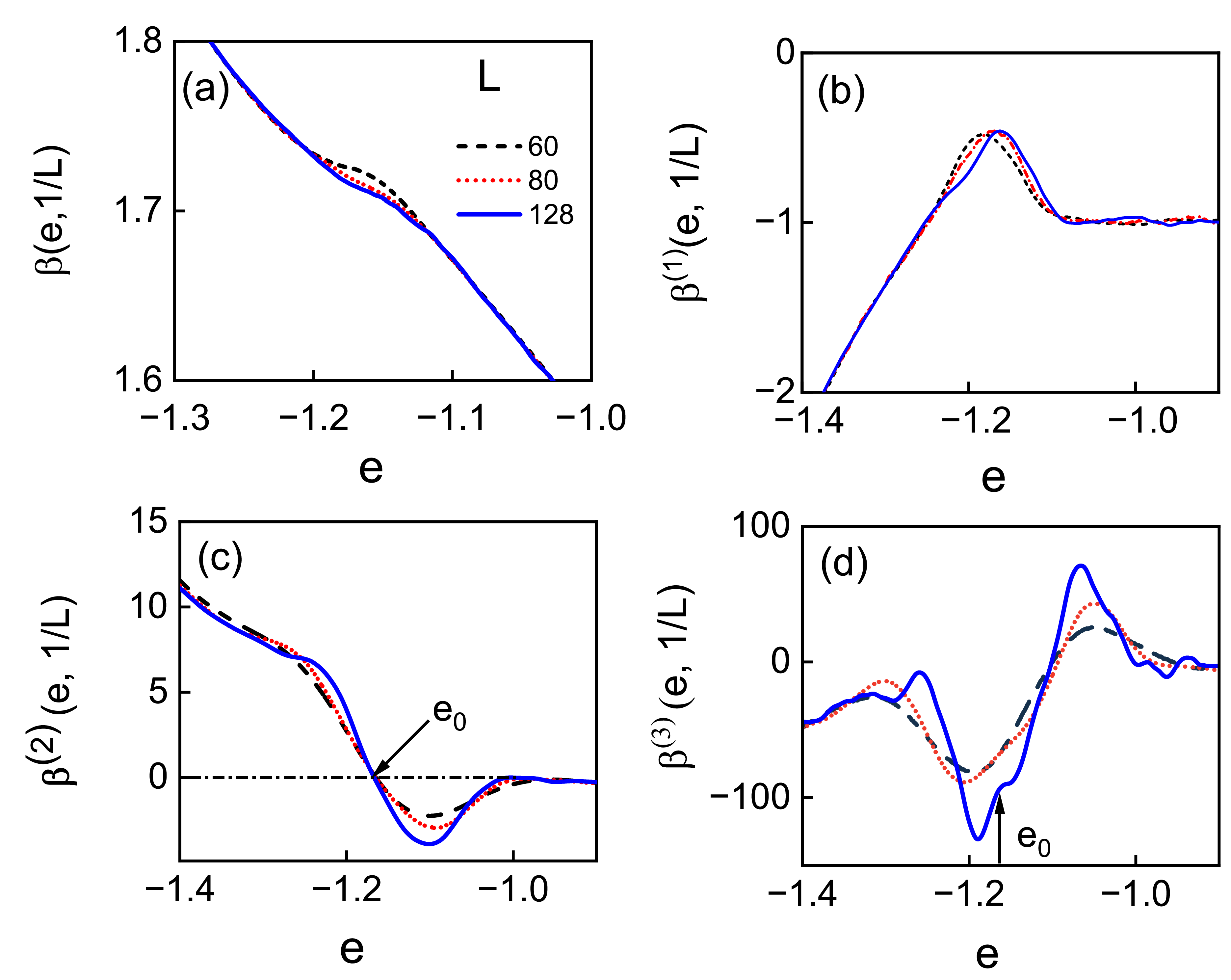}
\caption{(color online) {\it Energy derivatives of the micro-canonical entropy:} 
(a) The $\beta(e,1/L) \equiv d S(e,1/L)/de$ curve 
versus energy $e$(at micro-canonical bin level)  for sizes $L=60,80,128$. 
(b) The $\beta$ slope or $\beta^{(1)} (e,1/L)$ curve. 
(c) The $\beta$ curvature or  $\beta^{(2)}(e,1/L)$ curve. Curvatures 
vanish at points  $\{e_{0} (1/L)\}$ on  the dash-line. The marked 
$L=128$  inflexion point is  $e_{0} = -1.166$. (d) The third derivative 
$\beta^{(3)}(e, 1/L)$ curve.} 
\label{fig:7}
\end{figure} 
In constructing the Markov sequence of microstates, BMC and EAMC protocols
differ in their acceptance probabilities. For a given $\Delta E$ and $\Delta  S$
corresponding to changes in energy and entropy, respectively,
between the trial and initial states, the acceptances are given by 
$$P_{\text{acc}}^{\text{(BMC)}} = \textit {Min} [1, e^{-[\Delta E /T]}]$$
and 
$$P_{\text{acc}}^{\text{(EAMC)}} = \textit {Min} [1, e^{-[\Delta E /T -\Delta S]}].$$
These acceptance probabilities guide the progression of Markov chain.
The EAMC protocol depends explicitly on the entropy increment, while the 
BMC protocol does not: the Monte Carlo evolutions can qualitatively differ.

 The BMC acceptance probability depends only on energy increments 
 through the Boltzmann factor. The EAMC acceptance probability has the 
 same Boltzmann factor, but also has an entropy increment that can be written as 
 $\Delta S \equiv S (E+ \Delta E) - S(E)   
 =(\Delta E) \sum_{m=0}^{\infty}[\beta_{\text{eff}} ^{(m)} (e, 1/L)/(m+1)!] (\Delta E )^{m}$. 
 This is determined from the EAMC data by $\beta$-derivatives such as
$\beta_{\text{eff}} ^{(m)} (e, 1/L) = {d^{m +1} s(e, 1/L)}/ { de^{m+1}}$, from 
bin-wise variation of entropy $s_{\mu}$ with  energy $e_{\mu}$. 
The derivatives with $m=0,1,2,..$ are the entropy slope (or inverse 
effective temperature), the entropy curvature, the slope of entropy-curvature, etc. 
This expansion explains how in Fig.~\ref{fig:1}, the  BMC and EAMC protocols can
match at higher energies, but can differ at lower energies closer to entropy 
barriers, that only the latter can detect and traverse. The subscript 
is henceforth dropped for notational compactness, with derivatives 
written as $\beta^{(m)} (e, 1/L)$.

Apart from EAMC considerations, there has been much work based on 
energy derivatives of the microcanonical effective temperature, motivated 
by {\it energetically sharp} cooperativity changes in finite systems, such 
as nuclei, atomic clusters, or biomolecules \cite{Gross,Schnabel,Qi}. The 
derivatives identify signatures of the order of the phase transitions in finite 
systems from general considerations on signs of entropy derivatives and 
from matches to exact finite-size solutions of soluble models with first-order 
and second-order transitions. It might also be interesting to examine the 
exactly soluble ideal Bose gas, where there is an independent third order 
transition \cite{Bose}, with a global, zero wave-vector condensate, 
interpenetrated by the condensate thermal depletion.

The $\beta$ derivatives identify different independent phase transitions 
over system scales $\sim L$. Transitions can be preceded  and fore-shadowed, 
by softer and dependent, {\it precursor transformations} over finite scales 
$\ll L$ \cite{Schnabel,Qi}. The transformation and transition special 
energies $e =e_\text{tr}$, correspond to scale-dependent temperatures 
through the relation $T(e_\text{tr},1/L) \equiv 1/ \beta (e_\text{tr}, 1/L) = T_\text{tr} (1/L)$.  
Fig.~\ref{fig:7} plots $\beta(e,1/L)$ and its various derivatives versus energy 
$e$ for different $L$. 

First order transition signatures are positive slopes of $\beta (e, 1/L)$
from back-bending \cite{Schnabel,Qi}.  
Whereas Fig.~\ref{fig:7}(a) shows that  the decreasing $\beta (e, 1/L)$ curve flattens 
slightly, but has a negative slope through out. Thus $\beta$-derivatives of the 
model do not show first-order signatures. This is consistent with the zero 
latent heat and absence of hysteresis at $T_n$ in Fig.~\ref{fig:3} and 
Fig.~\ref{fig:5}. Further, there is smooth variation with temperature of the 
Binder energy cumulant (not shown here) \cite{Binder} negating first order 
behaviour. The four pieces of evidence clearly rule out a first-order transition.
 
 Second order transitions have first derivatives of $\beta$ with peaks that 
 shift with increasing size  to lower energies, while (negative) peak values
  rise to zero \cite{Schnabel,Qi}. Fig.~\ref{fig:7}(b) shows that for increasing 
  $L$, the model $\beta^{(1)}(e,1/L) (< 0)$ has peak positions that shift 
  towards higher energy values with increasing $L$, while peak values 
  themselves are insensitive to changes in $L$. Thus $\beta$-derivatives do 
  not show  second-order signatures. As mentioned earlier, $C_{v}$  (per site) 
  versus  $T$ curve is independent  of the size of the system, and its peak 
  does not sharpen in slope or asymptotically diverge, with increasing 
  system size \cite{Chiccoli,BKLPRL}. Further, the correlation length 
  (Fig.~\ref{fig:6}) is non-divergent \cite{BKLPRL}. The three pieces of  
  evidence clearly rule out a second-order transition.

We note that  derivative of the $C_{v}$ (per site) with 
$T$, $d C_{v} / dT \sim  -d^{3}f (T) / d T^{3}$, shows a jump, as in Fig.~\ref{fig:6}. 
This is direct evidence of  a  {\it third-order} transition in the Ehrenfest classification \cite{Callen}. 
With the transition order determined, it is interesting to examine for 
its manifestations in the higher order $\beta$-derivatives in this model. The 
energy curvatures $\beta^{(2)}(e,1/L)$ of Fig.~\ref{fig:7}(c) are negative for $-1.17 < e < -1.0$, 
and vanish at distinct (non-stationary)  inflexion points $\{e_0 (1/L)\}$ defined 
by  $\beta^{(2)} (e_{0}, 1/L) \equiv 0, \beta^{(1)} \neq 0$. For  $L=128$  the point $
e_{0} \simeq -1.166$  corresponds to $T(e_{0}, 1/L) \equiv 1/\beta(e_{0}, 1/L) = 0.585$, 
the  nematic transition temperature $T_{n}(1/ L)$. The independent 
third-order transition in our model is identified with a non-stationary point
 of inflexion in $\beta$, at points $\{e_0(1/L) \equiv e_n (1/L)\}$. Fig.~\ref{fig:7}(d) 
 shows that  for $L=128$ $ \beta^{(3)}(e,1/L)$ is flat at  $e_0$, or $\beta^{(4)} (e_{0}, 1/L)=0$.

 Just as for the Ising model \cite{Schnabel,Qi}, the $\beta$-curvatures for 
 different sizes shown in Fig.~\ref{fig:7}(c) curiously have a common 
 crossing point $e_\text{c}$ (not marked), reminiscent of  the Binder OP 
 cumulant \cite{BKLPRL,Binder}. The values at the crossing point are $
 e_{\text{c}} = - 1.174$, $\beta^{(2)}(e_{\text{c}}, 1/L) = 0.723$.

Finally, to check the separation of the precursor transformation and phase 
transition for large systems, a linear extrapolation  $1/L \rightarrow 0$ yields  
$T_{n}(1/L) \rightarrow 0.586$: the separation is small but nonzero, 
$(T_{p} - T_{n})/T_{n} \simeq 7 \times 10^{-3}$. 
  
\subsection{ Free-energy landscapes}
The entropy barrier crossings necessarily  involve correlated changes 
in the coexisting  nematic order parameter  $S_n$  and the unbound defect 
density $\rho_d$. Their  canonical free energy profiles could  be revealing.
\begin{figure}[h!]
\centering
\includegraphics[width=0.45\textwidth]{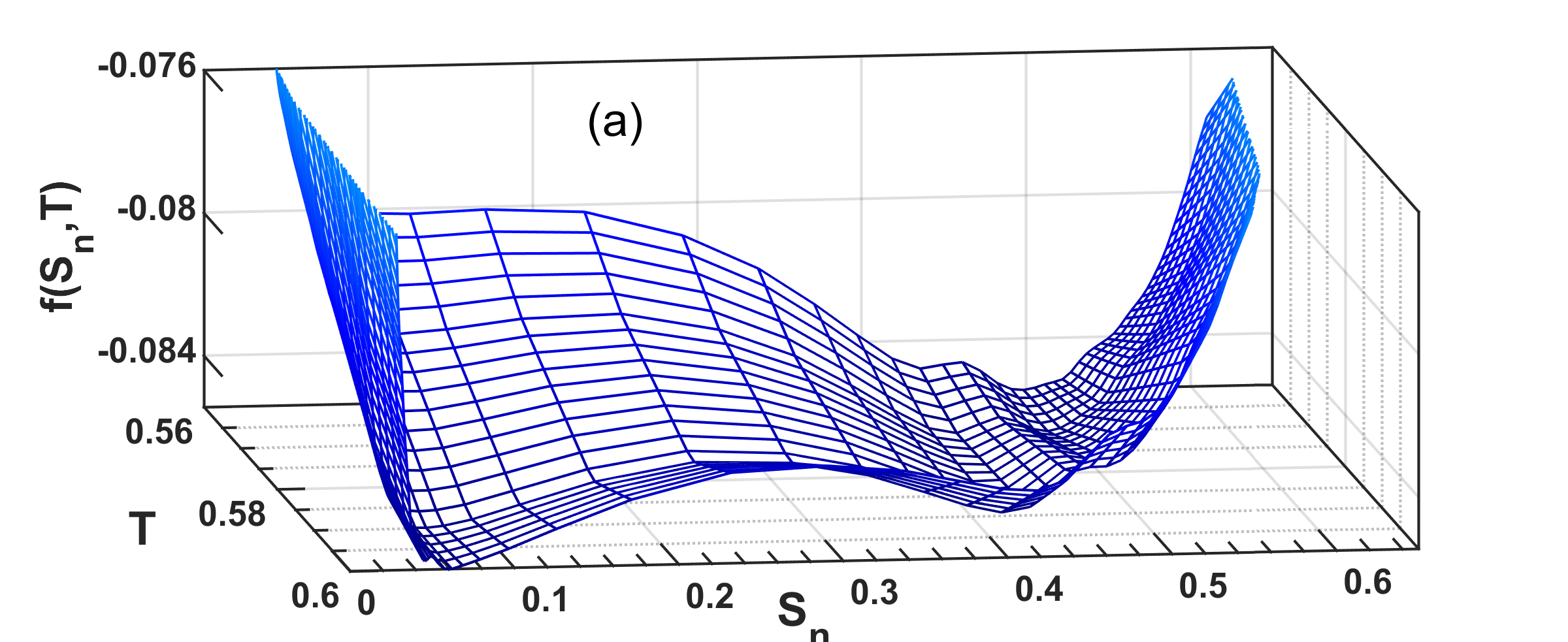}
\includegraphics[width=0.45\textwidth]{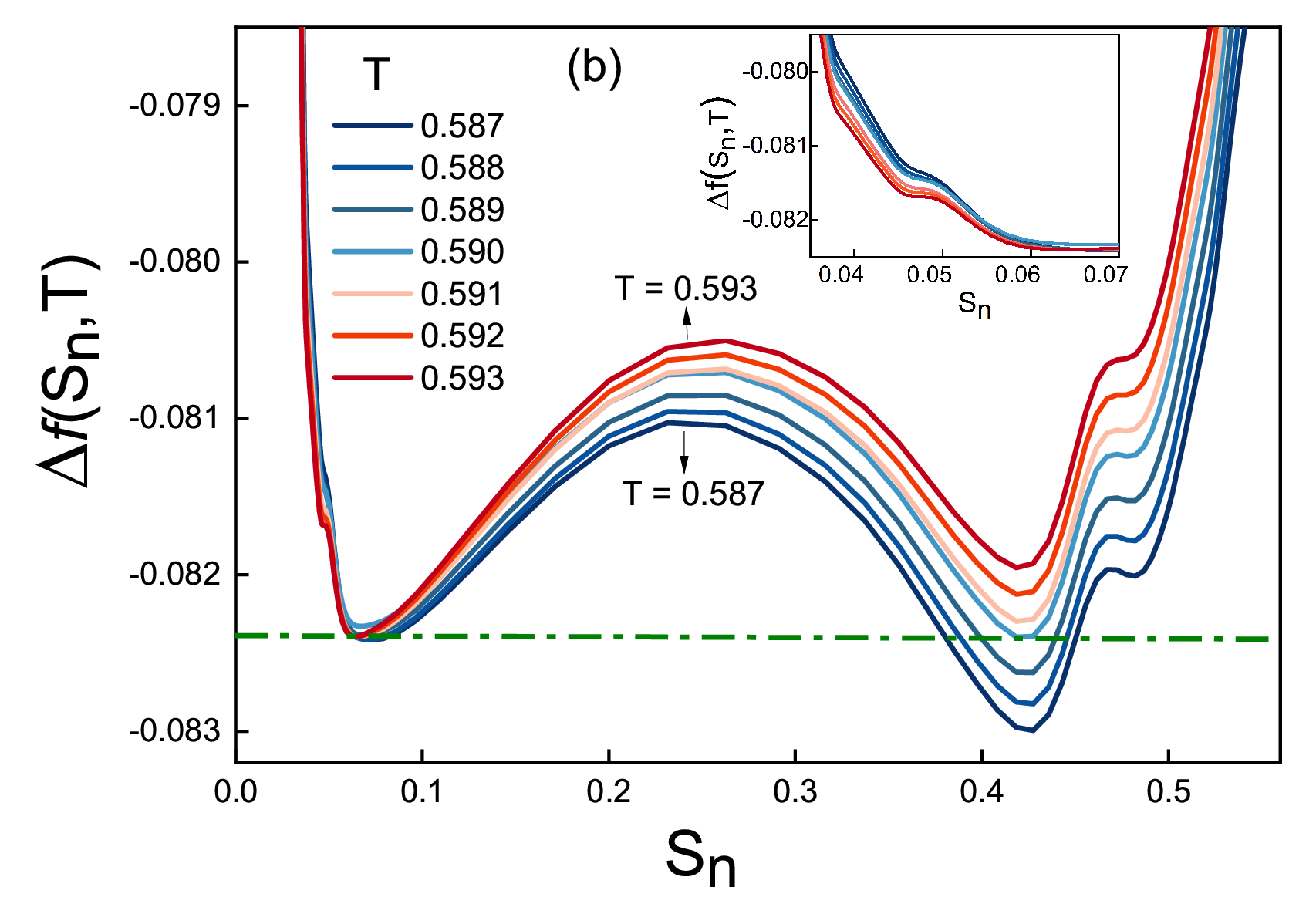}
\caption{(color online) {\it Free-energy landscape over the order parameter:} 
(a)  3D mesh plot of equilibrium  free energy density  
of the system $f(S_{n},T)$ over  the plane of equilibrium
nematic order  parameter $S_{n}$ and temperature $T$ for $L = 128$.
 (b) A background-subtracted plot of  $\Delta f(S_{n}, T) $ versus  $S_{n}$,
  has  a maximum from sparse configurations. The isotherms
 stack with smaller $T$ curves on the bottom. The $T=T_p =0.590$ isotherm
 tracks the bottleneck passage without latent heat, of transitional 
 configurations leading to increased order (left to right). Inset: 
 High temperature minimum at small $S_{n}$.}
  \label{fig:8}
\end{figure}    
Fig.~\ref{fig:8}(a) shows the  equilibrium free energy $ f(S_{n},T)$ 
as a 3D-mesh over order parameter and temperature in the transition region: 
it has an unusual shape of a `tilted washboard' potential. Compare the 
Fig.~\ref{fig:1} density of microstates over order parameter and energy.
Fig.~\ref{fig:8}(b) shows $\Delta f(S_{n},T) \equiv f(S_{n},T) - f(S_{n} =0, T)$.
The curves versus $S_{n}$ at constant $T$ are determined  from equilibrium  
EAMC data (and are \text{not} variational). The novel nematic free energies 
have a small $S_{n}$ minimum for isotropic clusters, a larger $S_{n}$ minimum 
for nematic clusters, and an  intervening maximum from sparser bottleneck states. For 
 $T=T_p= 0.590$   the finite-size director clusters around defect 
 cores have a rare passage  (without latent heat)  from small to large $S_n$, 
 with the same free energy at bottleneck entry, and exit.   
 For $T$ decreasing  through $T_{p}$ the emerging  `para-nematic' 
microstructures make a downhill run,  towards a final nematic phase at $T_n$.

 \begin{figure}[h!]
\centering
\includegraphics[width=0.45\textwidth]{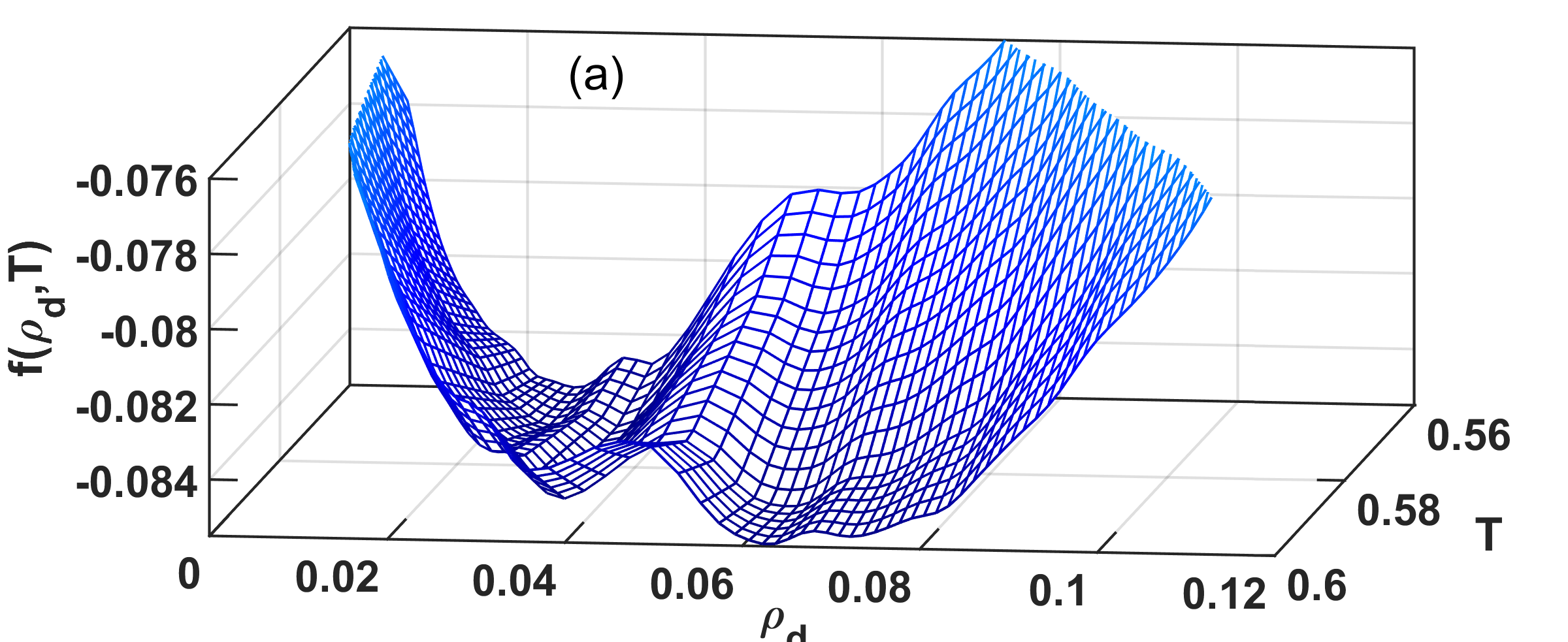}
\includegraphics[width=0.45\textwidth]{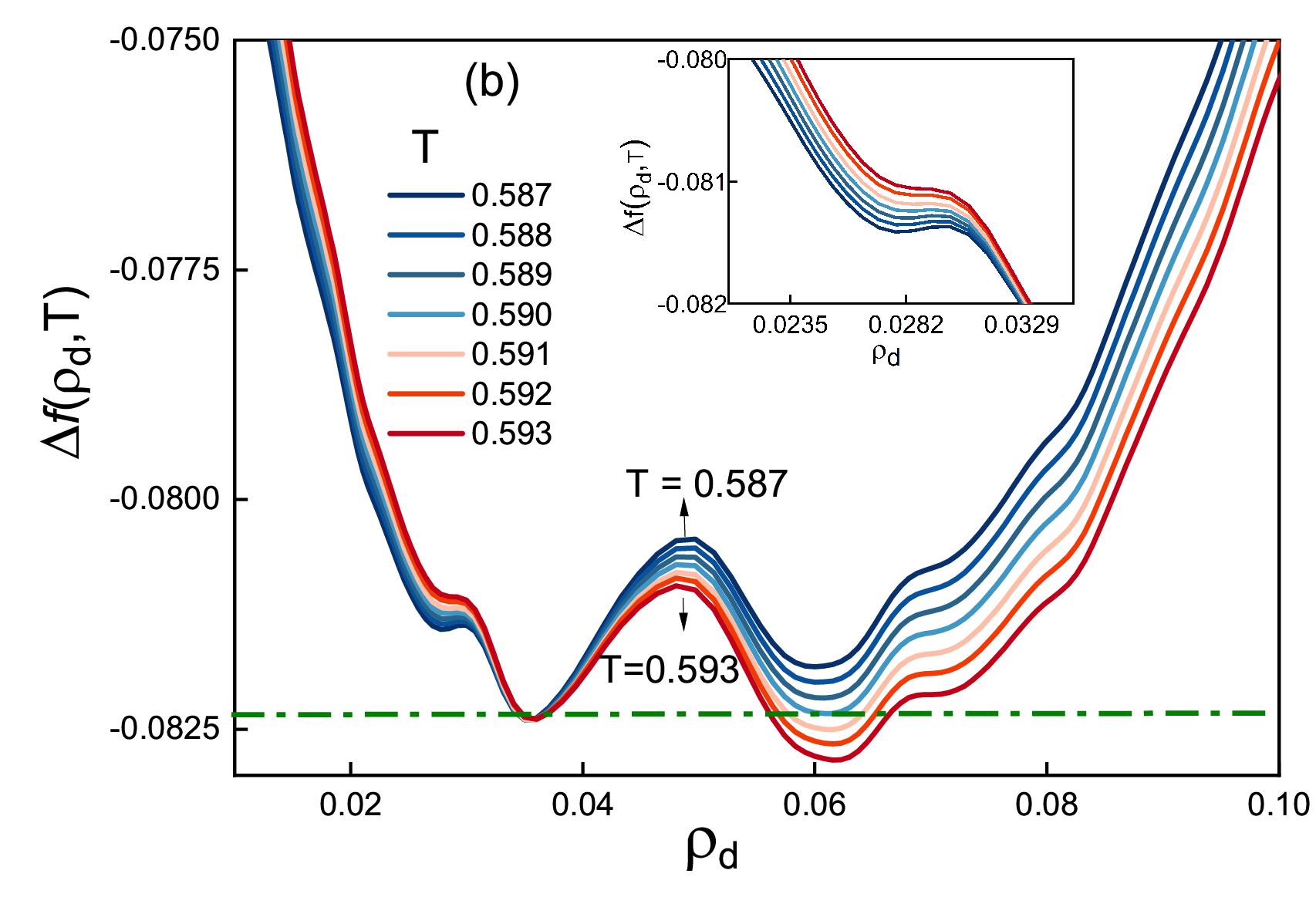}
\caption{(color online) {\it Free-energy landscape over the defect density:} (a)  3D mesh plot 
of equilibrium  free energy density  of the system $f(\rho_{d},T)$ over  the 
plane of equilibrium defect density $\rho_{d}$ or `disorder parameter', and temperature $T$.
 (b) A background-subtracted curve $\Delta f(\rho_{d}, T) $ versus  $\rho_{d}$, has a maximum 
 from  sparse  configurations. The isotherms stack with smaller $T$ curves on the  top. 
 The $T=T_p =0.590$ isotherm tracks the bottleneck  passage without latent heat, of transitional configurations leading to decreased disorder (right to left).
Inset: Low temperature minimum at small $\rho_{d}$.}
 \label{fig:9}
\end{figure} 
Figs.~\ref{fig:9}(a) and (b)  show  similar complementary free energy 
curves at different $T$, for $f(\rho_d, T)$ and $\Delta f(\rho_d, T)$. 
  
 \begin{figure} [h!] 
\centering
\includegraphics[width=0.45\textwidth]{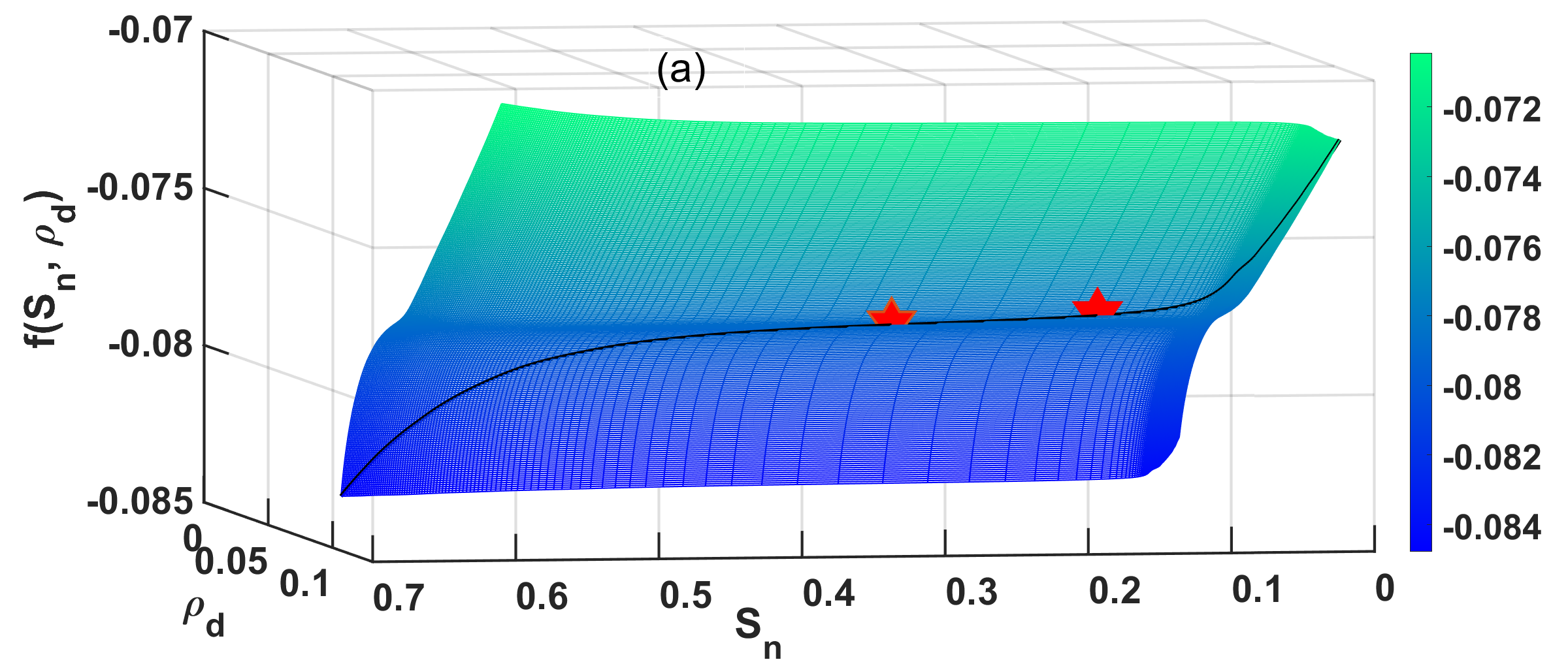}
\includegraphics[width=0.45\textwidth]{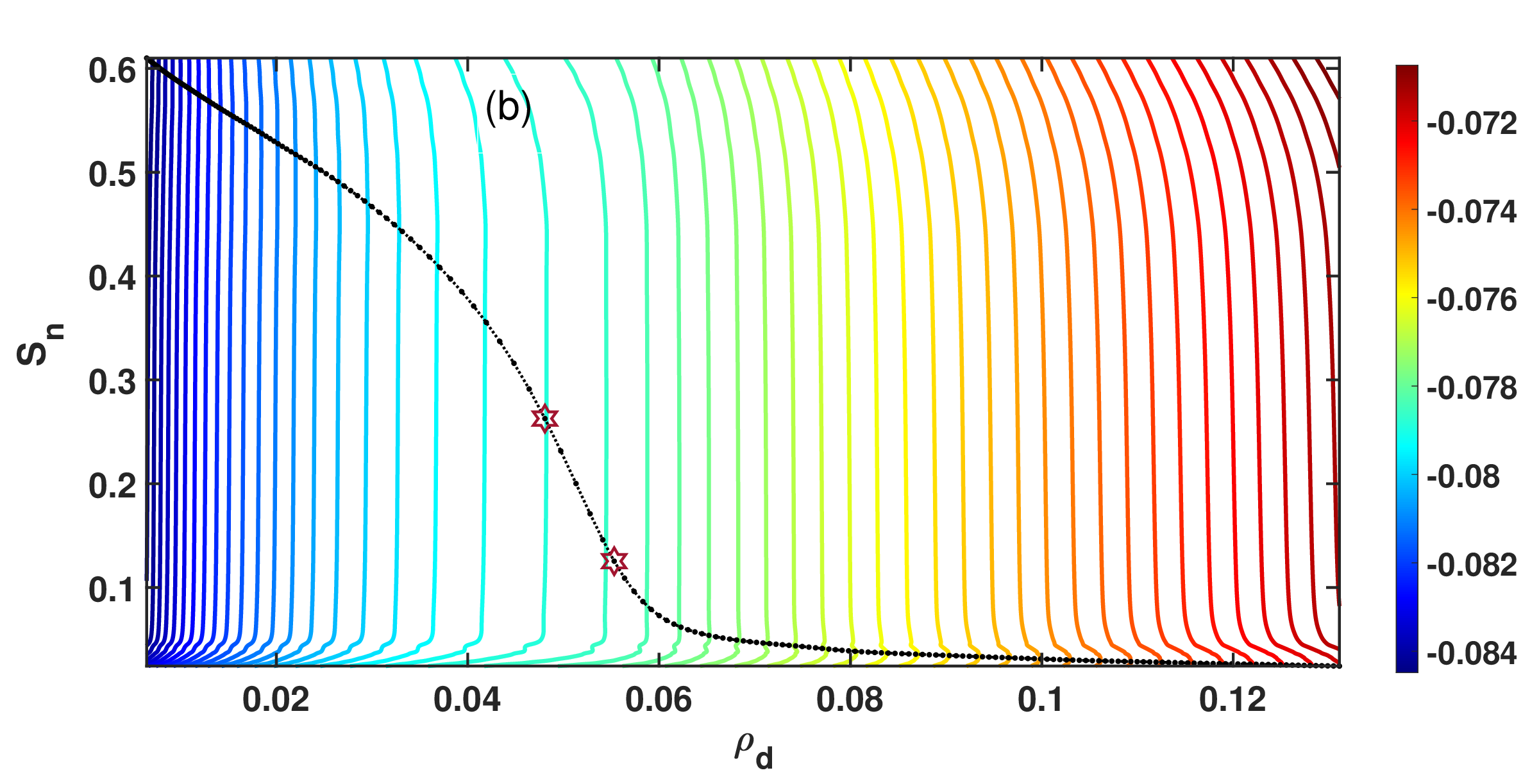}
\caption{(color online) {\it Free energy landscape over the order and disorder parameters:}   (a) 3D 
mesh plot of equilibrium free energy density of the system $f(S_{n}, \rho_{d})$ 
over  the $(S_n, \rho_{d})$ plane of the equilibrium coexisting order
($S_{n}$) and disorder ($\rho_{d}$) parameters.  
 The equilibrium path  of $(S_{n}, \rho_{d})$ on the free energy surface is 
 traced as a solid (black) line.  The star symbols denote ($S_n, \rho_d$) values at 
$T= T_{p}$ and $T_{n}$. (b) The background-subtracted  
free energy $\Delta f (S_n, \rho_d)$ has contours of constant value that 
show the range of values $S_n$ versus $\rho_d$ of the unbound defects 
in the nematic background, for the novel nematic. Superimposed are black
solid line of canonical equilibrium averages. The star symbols mark  
 ($S_n, \rho_d$) values at $T = T_{p}$ and $T_{n}$.} 
 \label{fig:10} 
\end{figure} 
The correlated and complementary  variation of order and disorder 
parameters is demonstrated by plotting the free energy
as function of $S_{n}$ and $\rho_{d}$.  In the mesh diagram of Fig.~\ref{fig:10}(a), 
smaller values of $\rho_{d}$  are associated with large values of $S_{n}$  
and vice versa. The solid line marks a rippled line  on this flat free energy 
surface that separates almost flat convex and concave segments. Along the
 `line of inflexion'  on the surface, the  points ($S_{n}, \rho_{d}$) at 
 $T= T_{p}$ and $T_{n}$, are marked by stars.

The contour projection of these data onto the plane of the equilibrium 
variable pair ($S_n, \rho_d$) in Fig.~\ref{fig:10}(b), shows the entropy 
barrier region has contour lines that are almost parallel, and almost 
equally spaced. This reflects the flatness of the free energy surface, 
related to the softness of the (third-order) transition. The solid line 
corresponds to equilibrium variation of $S_n$ with $\rho_d$, with 
the marked points at $T_{p}$ and $T_{n}$ indicating the bottleneck 
entry and exit. The bottleneck region shows a large increase of 
$\sim +0.2$ in $S_{n}$ for a small decrease of $\sim -0.02$ in $\rho_{d}$. 
  
\begin{figure}[h!]
\centering
\includegraphics[width=0.45\textwidth]{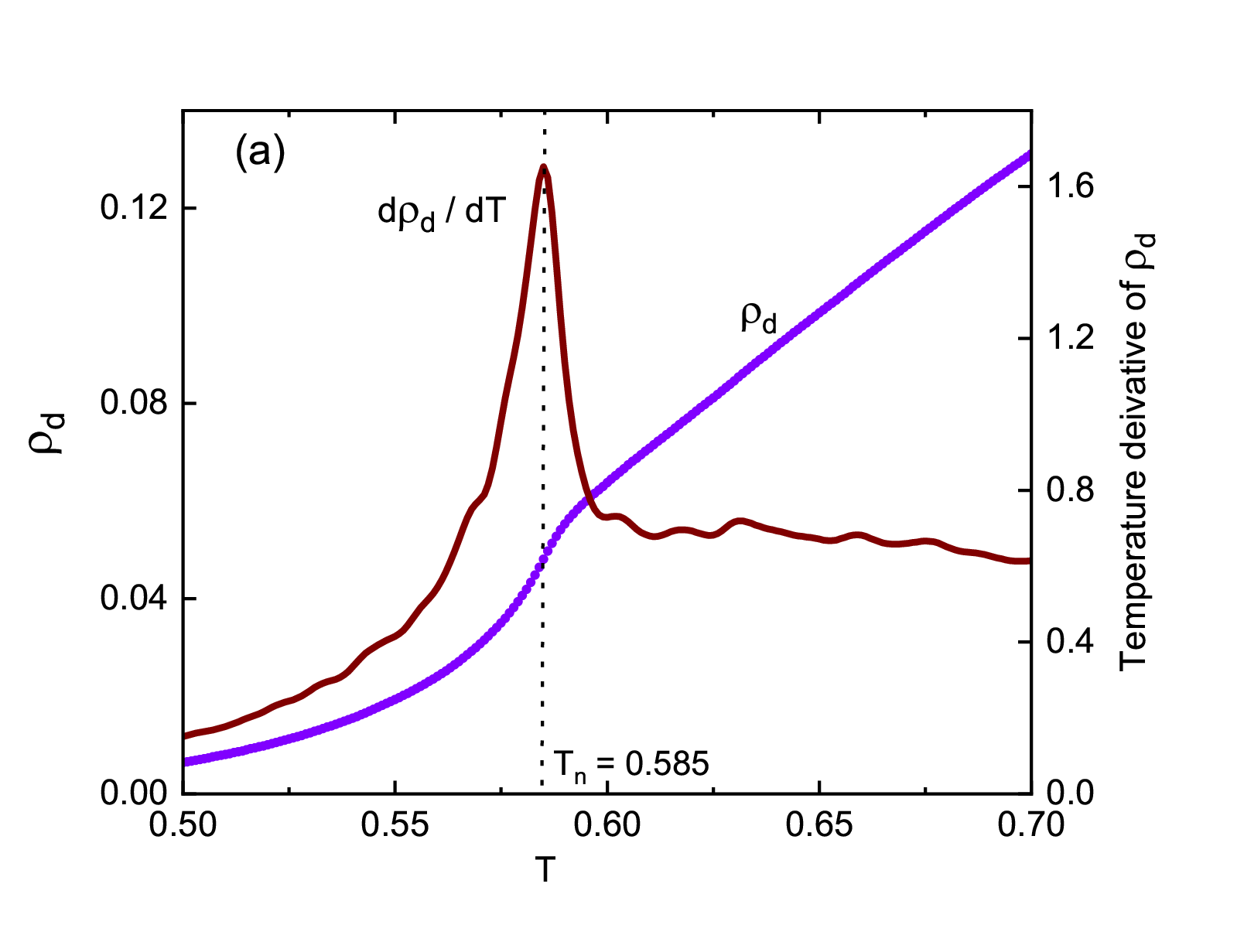}
\includegraphics[width=0.45\textwidth]{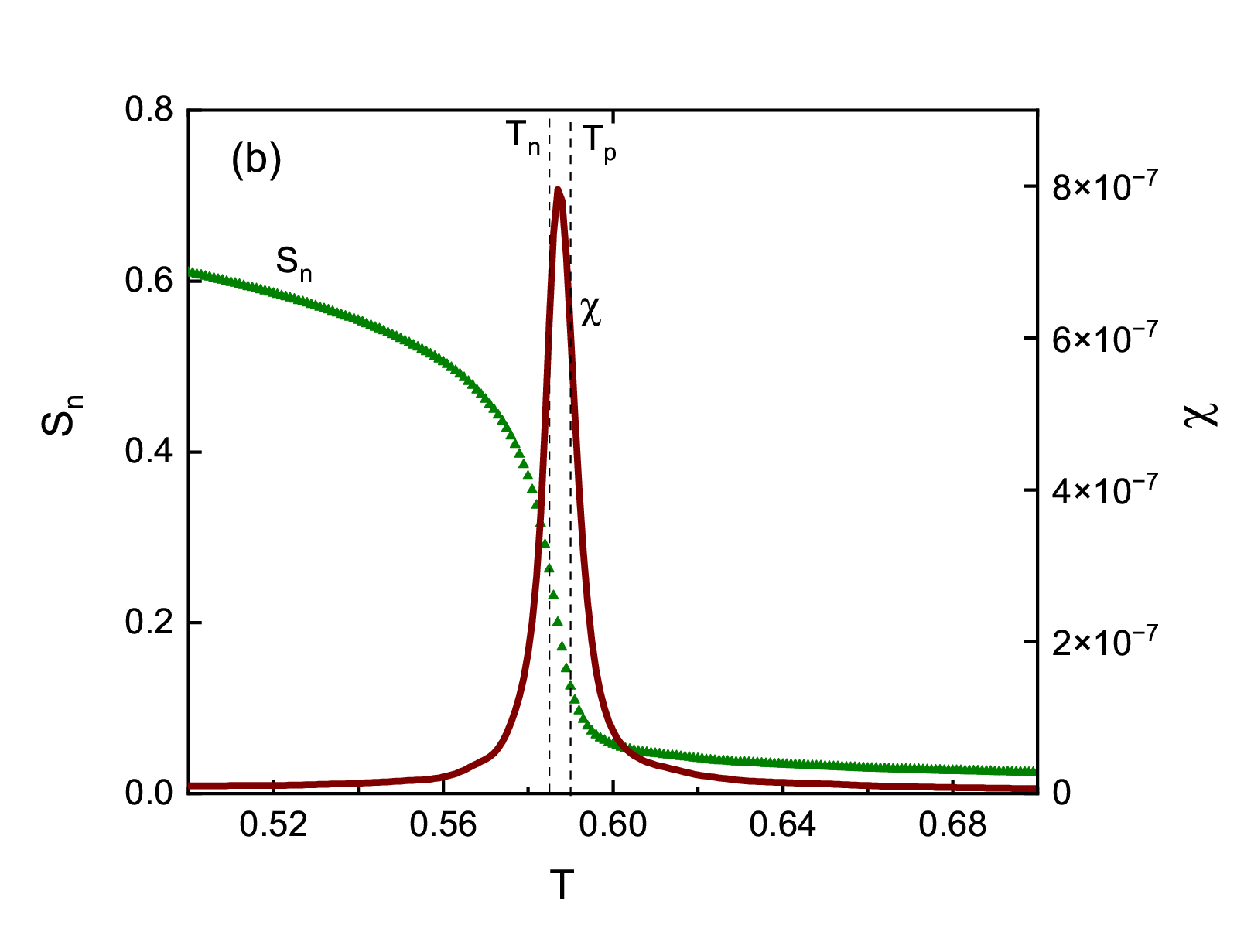}
\caption{(color online) {\it Singularities of $\rho_d$ and $\chi$:}
 (a) Equilibrium  defect density $\rho_d (T) $ and its temperature derivative versus 
 $T$, that shows a cusp at $T_{n}$. (b) The equilibrium  nematic order $S_n$ and its 
 nematic susceptibility $\chi (T)$ versus $T$, that shows a cusp at a 
 $T=T_b=0.587$, just above $T_{n}$ and inside the  bottleneck 
 window $ (T_{p} -T_{n})$. See text. }
\label{fig:11}
\end{figure} 

Fig.~\ref{fig:11} presents the temperature variation of equilibrium $\rho_{d}$ and 
$S_{n}$ in the neighbourhood of the transition. In Fig.~\ref{fig:11}(a),
$\rho_{d}$ is seen to have a smooth variation across the transition, 
whereas its temperature derivative has a cusp  at $T_{n}$ = 0.585, 
coincident with the specific heat cusp. In Fig.~\ref{fig:11}(b), the equilibrium
variation of $S_{n}$ and the nematic susceptibility $\chi$ are plotted
as  functions of temperature. $S_{n}$ has a steep but smooth rise on cooling, and 
$\chi$ has a cusp at a $T=T_b$ just above transition, and inside the 
bottleneck window $T_{p} - T_{n}$. This suggests that in passing through the 
entropy barrier, there is a rare re-arrangement of director clusters
 from random orientations in the isotropic phase, to a locally tilted dressing of defects. 
 								
\subsection{Transformation transients} 
 \begin{figure}[h!]
\centering
\includegraphics[width=0.5\textwidth]{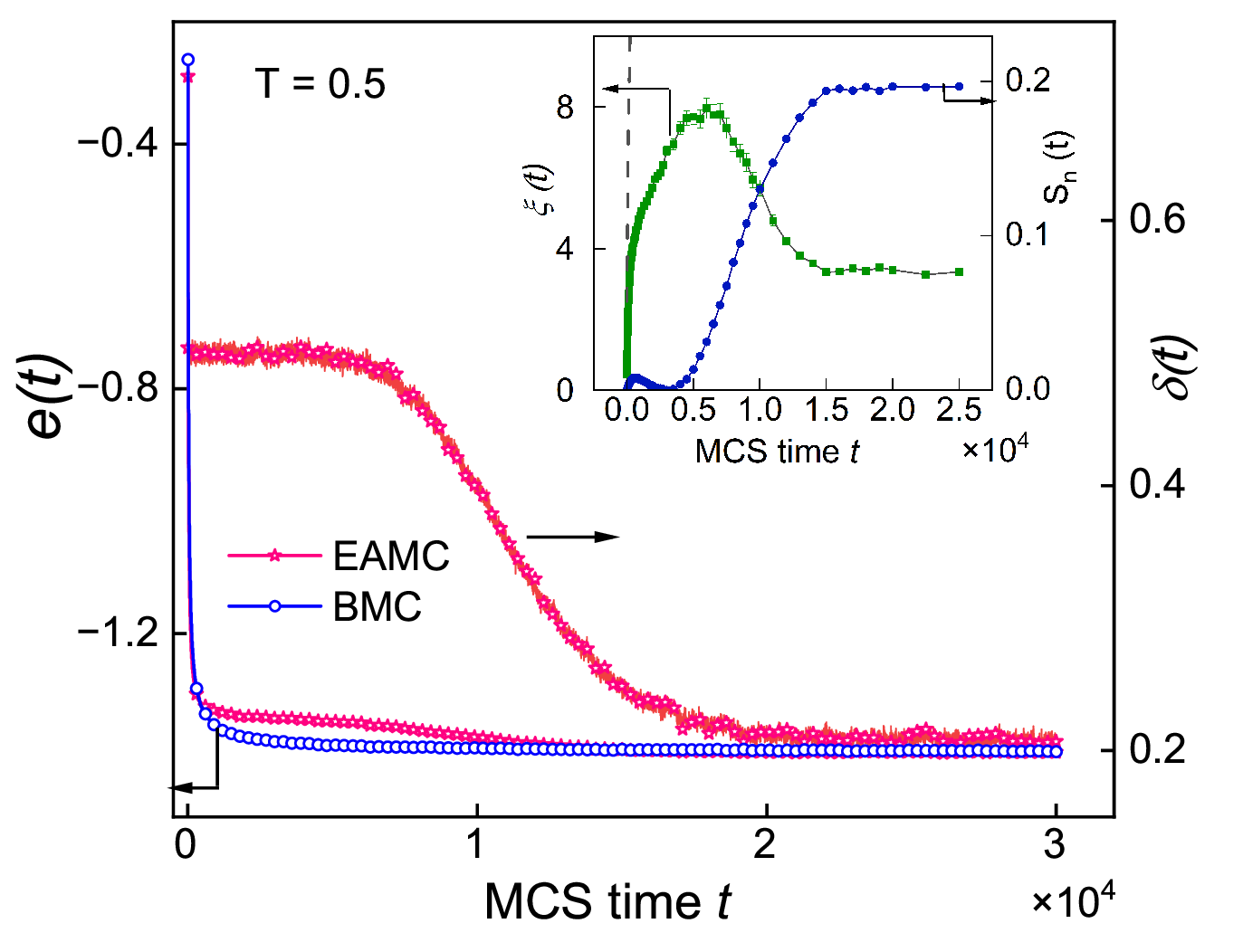}
\caption{(color online) {\it System evolutions in MC time:} After a quench to $T=0.5$,
instantaneous values of relevant observables are tracked as a function 
of Monte Carlo time ($t$). Main figure shows 
free-running dynamics of energy $e(t)$ under BMC and EAMC protocols, 
with re-equilibration after $t \approx t_X = 1 \times 10^{4}$ MCS.  The EAMC 
evolution of topological parameter $\delta(t)$ is also shown. Inset: The
the order parameter $S_{n} (t)$, and correlation length  $ \xi (t)$ 
initially evolve on expected lines, with the latter locking to the   
  BKT coarsening length $L_{c}(t)$ (dashed line). Both deviate from 
  these paths, but differently, at the onset of the re-equilibration regime. 
  Development of a large $\xi (t)$   during the long search period facilitates 
  finding  a symmetry-breaking PES pathway to the novel nematic equilibrium phase.}
\label{fig:12}
\end{figure} 

\begin{figure}  [h!]
\centering
\includegraphics[width=0.5\textwidth]{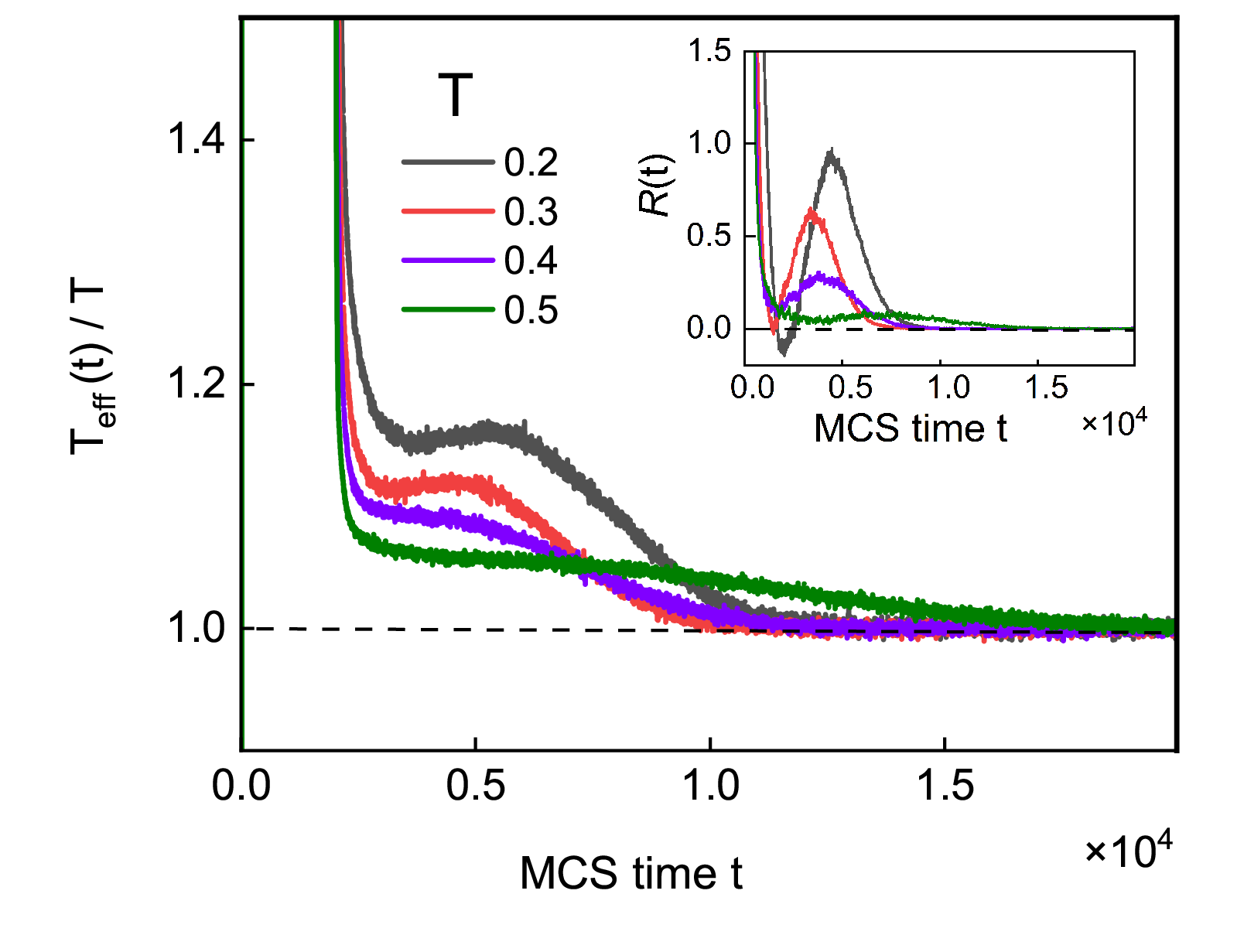}
\caption{(color online) {\it Correct re-equilibrations in MC time:} Profiles 
of $T_{\text{eff}} (t)$, showing (re-entrant) cooling 
and  final correct re-equilibration beyond $ t\approx  t_X$ to 
quenched bath temperatures $T$.  Inset: The (per-site) 
total entropy production rate $R(t)$ from 
entropy-barrier crossings, correctly re-equilibrates to zero.}
\label{fig:13}
\end{figure}  
 To explore non-equilibrium pathways in the configurational 
space during the  defect core and nematic  cluster co-evolutions, we  
implement a free-running PES-type dynamics \cite{Ritort,Bonilla}  with the EAMC quench 
protocol. The system is initially equilibrated at 
 $T_{\text{init}}= 2.0$, and allowed to 
evolve for time $t$ up to  $ 5  \times 10^{3}$ Monte Carlo  sweeps (MCS). 
The bath temperature is then suddenly quenched at $t = 0$ to fixed temperatures 
$T = 0.2, 0.3, 0.4, 0.5$,  and 2000  independent quench runs are simulated 
to generate a representative ensemble of pathways. Averaged correlation functions  
$G(r;t)$ are calculated at 75 chosen time points $t$  distributed over
$ 2.5  \times 10^{4}$ MCS till equilibrium is reached during each  
evolution. As for the static case \cite{BKLPRL}, the EAMC-derived 
$G(r,t) $ is parametrized as $G(r,t)  = G(0, 0)  \exp(-r / \xi(t)) + S_{n}^{2}(t)$, 
where the initial $S_{n} (0)$ was found to be practically  zero at 
$T_{\text{init}}$. The run-averaged  transient variables 
$e(t), S_{n} (t),  \delta(t), s(t), \beta_{\text{eff}} (t) \equiv \beta_{\text{eff}} (e(t))$ are 
computed. The quenches possibly generate the statistics of  `large deviations' 
from equilibrium \cite{SatyaICTS,Touchette}, that should vanish on correct re-equilibration.

 The main Fig.~\ref{fig:12}  shows for a quench to $T = 0.5 < T_{n} < T_{p}$, that 
 the EAMC evolution of energy $e(t)$ deviates from  the BMC curve, detouring 
 through a flattening regime, before  equilibration  beyond 
 $t \approx t_{X} = 1 \times 10^{4}$ MCS to a value nearly 
 energetically degenerate with the BMC data.  The post-quench topological parameter
 $\delta (t)$ remains flat over this time scale $t_{X}$. Whereas  $S_{n}(t)$ and  $\xi(t)$ promptly initiate complex and  correlated evolutions.

 The inset of Fig.~\ref{fig:12}  shows that  $S_{n} (t)$ rises slightly from 
 its $t=0$ value of zero, and falls back to nearly zero at 
 $ t \sim 0.3 \times 10^{4}$ MCS. The  evolving correlation length $\xi (t)$  
 is locked to the BKT coarsening length: $\xi(t) = L_{c} (t) \sim [t / \ln t] ^ {1/2}$ 
 \cite{Rojas, Dutta05, Singh, Ozeki}, until $\xi (t) \sim 3.6$ 
 ($t \leq 300  MCS$). The correlation  function  
 $[G(r, t) - S_{n} (t)^{2}] / G(0, 0)$ in this locked regime exhibits 
 dynamical scaling, or data collapse  in $r / \xi(t)$ (not shown here). 
 Thereafter, the  transient nematic cluster sizes rise to $\xi (t) \sim 8$ lattice units, 
 before falling to an equilibrium value $\xi_{n} (T) \sim 3.4$ after 
 the a comparable relaxation time $t \approx t_{X}$ MCS. 
There is thus a temporal crossover from a regime of dynamical 
coarsening without symmetry breaking, to a regime of  a static 
correlation with symmetry breaking. The evolution discussed above is reminiscent
of a generic partial equlibrium scenario (PES). It governs the sequential  
 passage of a non-equilibrium  system between micro-canonical shells of 
 decreasing energy, through inter-connecting bottlenecks. These PES evolution 
 ideas  were earlier applied (using BMC protocols) to entropy-barrier 
 passage of ageing harmonic oscillators, and to re-equilibration  of martensitic 
steels \cite{Ritort,Bonilla,Garriga,Crisanti04,Crisanti13,ShankarEPL}.  

 Fig.~\ref{fig:13} confirms  that the $T_{\text{eff}} (t)$ and   total entropy
production rate $R(t)$ find  correct re-equilibrations to $T$ and zero, 
respectively at times $t \approx t_{X}$. Here, the total entropy change of a system-plus-bath, 
$ dS_{\text{total}} = dS(E) + dS_{\text{bath}} (E_{\text{bath}}) > 0$ at constant 
$E + E_{\text{bath}}$, yields an entropy production rate  in intuitive form,
$R(t) \equiv {\dot S}_{\text{total}}/ N  = [\beta_{\text{eff}}(t) - \beta]~ {\dot e}(t) $. 
Time-averaging a quantity  like  $q(t,T) \equiv [1 - e^{-t R(t)}]$ \cite{SatyaICTS,Touchette}
 for the different quenches $T$ of  Fig.~\ref{fig:13}  yields values 
 $(5.2, 2.6, 1.8, 1.5) \times 10^{-3}\ll 1 $, implying that in the Inset of Fig 13, the entropy production rate $R(t)$ falls asymptotically to zero faster than $\sim 1/t^2$ during 
  the re-equilibration.
 
\subsection{ Condensation to special angles} 
The presence of defects of a fixed topological charge could constrain 
permissible directions of the nematic director, with respect to the 
laboratory system (with the $2D$ lattice of the model defining the equatorial
 $XY$-plane, say). As described earlier, free-running evolutions after a 
 quench are probed by EAMC-based quench simulations to selected bath
temperatures in the neighbourhood of $T_{n}$. During the equilibrated 
evolution at a chosen temperature, a microstate is randomly chosen and 
direction of its nematic director is computed by determining the normalized 
eigenvector corresponding to the (absolute) maximum eigen value of the 
ordering tensor. The direction cosines of this vector with respect to the 
laboratory frame correspond to realization of the director orientation  associated with a representative member of the equilibrium ensemble at that temperature. During equilibrium
evolution at each temperature, $5 \times 10^{3}$ microstates were 
sampled at intervals of $10 \times 10^{3}$ MCS. Data on the direction 
cosines $\{\alpha_k \}$ with $k =1, 2, 3$
of these microstate directors, are recorded. They are graphically
 represented as a distribution of points on the surface of a unit sphere.
 
 \begin{figure} [h!]
\centering
\includegraphics[width=0.5\textwidth]{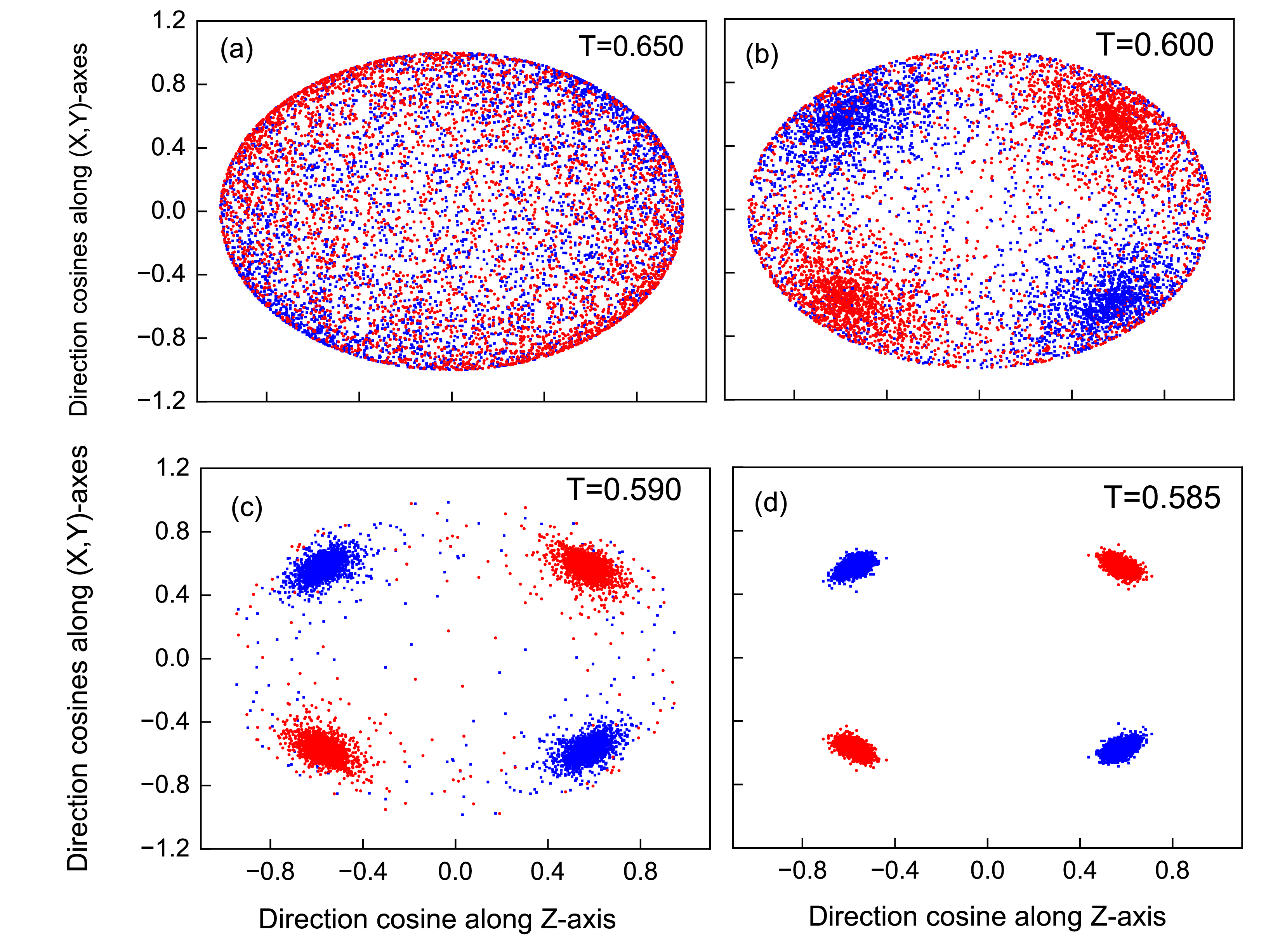}
\caption{(color online) {\it Condensation to special angles:} The 
four-panel (Z,X=Y)  plots depict the progressive macroscopic occupation of the 
orientational distribution of $n=3$ equilibrium  directors, at direction cosines $\{(\cos \theta_i, \cos \phi_i)\}$ that condense towards macroscopic occupancies of 
 $(\pm 1/ \sqrt{3}, \pm 1/\sqrt{2})$ at temperatures bracketing the nematic transition : 
 (a) $T= 0.650$, (b) $T= 0.600$, (c) $T= T_{p} = 0.590$, (d) $T= T_{n} =0.585$.
}
\label{fig:14}
\end{figure}
    
 Fig.~\ref{fig:14} depicts the distributions as XYY plots, at four temperatures 
 $T$ = 0.650, 0.600, 0.590 and 0.585, in the region of the crossover. In 
 each of these four panels, the direction cosines $\alpha_1, \alpha_2, \alpha_3$  
 are plotted with $\alpha_1$ along the horizontal axis and $\alpha_2$ ($\alpha_3$) along the right (left) vertical axis.  For $T=0.650$ the distributions of representative directors in the 
 canonical ensemble are random over the unit sphere.
 As the ($n=3$) system is cooled, the microstate directors  begin to 
 prefer direction cosines $\alpha_1, \alpha_2, \alpha_3$ of values $\pm 1/\sqrt{3}$. 
 
 Fixing a laboratory frame as where the 2D lattice system defines the 
 equatorial plane, and with a polar axis  perpendicular to it, the preferred 
 physical orientation of the ($n=3$) directors will be along one of the four 
 body diagonals. So their  preferred polar angle 
 projections  compare to $\{\cos \theta_i = \pm 1/\sqrt{3}\}$,  the presence of both 
 signs reflecting $Z_2$ symmetry. The projections on to the equatorial plane 
 will bisect the angle between the normal axes of the square base, so the 
 preferred azimuthal angle projections will be $\{\cos \phi_i = \pm 1/\sqrt{2}\}$. 
 This is consistent with the presence in the equatorial plane of  defects of  
 topological charge $1/2$, that for loops enclosing the core,  enforce 
 azimuthal increments $\pi/4$.
 
 The distribution data show all of the expected features. Fig.~\ref{fig:14}(b) for 
 $T=0.600$ reveals that preferred angles emerge. Fig.~\ref{fig:14}(c)  for $T= 0.590$ 
 and  Fig.~\ref{fig:14}(d) for $T=0.585$ show a condensation, and sharpening 
 to macroscopic occupation, of preferred angles. The complex onset of the 
 novel nematic could be clarified by simulations on cooling, of the direct and 
 cross correlations of polar and azimuthal angles in both coordinate and 
 Fourier space, examining behaviour at both large and small separations and wave vectors.

An expansion of the Hamiltonian of Eqn.~(\ref{eqn:1}) in spherical harmonics
 provides a possible scenario for the onset of ordering to a  novel nematic. 
 
For  ($ n=2$) the 2DLL model is
\begin{equation}
H (\{\phi_{ij}\}) = - \epsilon{\sum_{<ij>} }\{ 1+ \frac{3}{4} \cos(2\phi_{i j})-1]\}
\label{eqn:2}
\end{equation}
is dependent on azimuthal angles $2\pi > \phi_i \geq 0$, but only 
through differences $\phi_{ij} \equiv \phi_{i} - \phi_{j} $ across lattice bonds.  
The fluctuations suppress long range azimuthal order \cite{MWH1,MWH2,MWH3,Rice}, 
and induce azimuthal defects. Fields coupled to defect cores in 2D can affect 
such transitions \cite{Jenkins,Palle,Marino,Pearl}.
 
For the $(n=3)$, 2D LL model of Eqn. (\ref{eqn:1}), a spherical harmonic 
expansion \cite{expansion} yields
\begin{equation}
H (\{z_i, z_j,\phi_{ij} \}) =  -{\frac{3\epsilon}{4}} \sum_{<ij>} \sum_{m = 0,1,2} B_m (z_i , z_j)  \cos(m \phi_{ij})
\label{eqn:3}
\end{equation} 
 where polar angles  $\pi > \theta_{i} \geq 0$ appear only through polar 
 angle projections  $z_i \equiv \cos \theta_i$  in the expansion coefficients, that 
 split into factors at sites,  $B_{m} \equiv b_m (z_i) b_m (z_j)$. The interaction 
 resembles the `generalized XY' model, that can have ordered phases \cite{Romano,daSilva}. \\ 
 
 For the $m=0$ azimuthally isotropic term, the coefficient factor is 
 $b_0 (z_i) = \sqrt{3} ({z_i}^2 -1/3)$, while the  $m=1, 2$ azimuthally anisotropic 
 terms have 
 factors $b_1 (z_i) =2 z_i \sqrt{(1-{z_i}^2)}$ and $b_2 (z_i) = (1- {z_i}^2)$.
 
 For $z_i =1$ or directors only along the polar axis with no equatorial plane 
 components, the ground state energy $\sim -\epsilon$ is recovered.
For $z_i =0$ or directors only in plane, the ($n=2$) Hamiltonian of Eqn.(\ref{eqn:2}) is recovered.
For $z_i = \pm 1/\sqrt{3}$ or director tilts macroscopically occupying an off 
plane, `magic' angle value,  the $m=0$ azimuthally symmetric term $B_0$ vanishes, 
leaving $m =1,2$ anisotropy.
 
Locally averaging polar projections of clusters around the defect cores, 
${z_i}\rightarrow <{z_i}> \equiv {\bar{z}}$ yield approximated coefficients 
$B_m ({\bar z})= {b_m (\bar z)}^2$ that sum to unity,  
$\sum_{m =0,1,2}(3/4) B_{m} (\bar z) =1$. Eqn.~(\ref{eqn:3}) then becomes  
\begin{equation}
\begin{split}
H  &= - \epsilon (1 + [(1-\frac{3}{4}(B_{0} (\bar z) 
  +B_{1}(\bar z))] [\cos 2 \phi_{ij}-1] \\
&  + [\frac{3}{4} B_{1}(\bar z)][\cos \phi_{ij}-1]) .
\end{split}
\label{eqn:4}
\end{equation} 

At high temperatures, azimuthal angles fluctuate freely due to 
azimuthal spin waves, and polar angles fluctuate about the 
equatorial plane with zero average out-plane tilts ${\bar z}(T) =0$. 
The azimuthal  defects with bare core fluctuations  yield the 
broad range over all angles of Figs.~\ref{fig:14}(a) and (b). As 
temperatures are lowered, dressed cores can have clusters that 
fluctuate around nonzero out-plane tilts, ${\bar z} (T) \neq 0$.
   
A  {\it  conjectured } cooling scenario is as follows.
i) For $T \gg T_p$, the isotropic phase has azimuthal angle defects, 
with only `bare' cores. The separation between unbound bare cores 
defines \cite{Kosterlitz}  the BKT length $\xi_{+} (T)$. Nematic 
fluctuations can occur in general on a scale of the equilibrium 
nematic correlation length $\xi_n (T)$. For $T \geq T_{p}$,  there 
can be a `band' of out-plane polar angle fluctuations, between two bare 
cores. The two physically distinct lengths are locked and equal, 
$\xi_{n} (T) = \xi_{+} (T)$. On cooling, the sharply rising singular 
BKT length forces the band to stretch. At a transformation temperature 
$T= T_{p}$, the band splits to re-center, forming short-range nematic 
clusters around the cores of size $\xi_{n}(T) \leq \xi_{+}(T_{p})$. This 
downward peel-off leaves a sharp cusp in $\xi_{n}(T)$ at $T_{p}$. Here, 
$C_v (T)$ has a softer structure of a curvature maximum at $T_{p}$. 
See Fig.~ \ref{fig:6}  and Fig.~ \ref{fig:2}\\
ii) For $ T_{p} \geq T >T_{n}$, clusters around defect cores can have 
polar projections that fluctuate around out-plane tilt values. The dressed 
(unbound) defects  of low density  $\rho_d (T) \ll 1$ form a dilute, itinerant, 
para-nematic fluid. Such short range polar order is not easily deleted by 
long wavelength azimuthal spin waves.

 The angular condensation commences at $T_{p}$, and completes with all 
 defects dressed, at a $T=T_{b}$, just above $T_{n}$. 
The randomly located dressed defects all have local polar angle tilts, that 
could effectively interact, through azimuthal correlations with an 
expected  power law decay. 
Cases of permissible ordering have been mentioned \cite{MWH3}. The local polar tilts at every dressed defect, could `de-localize', to form a global polar tilt of nematic order at $T_{n}$. See 
Figs. 8(b), 9(b) and 11.\\
iii) For $T_{n} \geq T >T_{\text{BKT}}$, the dressed defects in the 
global nematic background will progressively bind in $\pm 1/2$ pairs.  \\
iv) For  $ T_{\text{BKT}} \geq T >  0$, the fully paired 
defects will gradually annihilate on cooling, tending to a uniform nematic. 
On warming, dressed defects are nucleated in pairs, and the scenario repeats in reverse order.

Dual transforms \cite{Kosterlitz,Young,Savit,Shenoy90,Shenoy93,JJA} on the
 above Hamiltonian could yield defect-defect interactions mediated by
 azimuthal spin waves, as 
for Eqn.~(\ref{eqn:2}). An internal topological weight parameter, dependent 
on the dressed core  $\lambda(T) = \lambda ({\bar z}(T))$ could appear, with
 real-space Renormalization Group flows in the defect coupling and fugacity  
 plane \cite{Kosterlitz} now suitably modified \cite{Catterall, Ozeki}. 
Interestingly, an added {\it external} topological field parameter $\lambda_0$ 
on each defect core resulted in enhanced transition temperatures and long 
range order, for the $(n=3)$ 3D Heisenberg, the $(n=2)$ 3D planar rotor and the ($n=3$) 2D LL models \cite{Dutta04,Lau,Kohring}.
 
\section{ Summary and further work}

The  novel features of the configuration space of this model  were studied 
with the EAMC protocol. An energy-uniform random walk of the system bracketing 
the crossover energy region, uncovered the existence of an entropy barrier 
between the disordered and nematically ordered regions of configuration 
space, with the bottleneck starting at $T_{p}$ and ending just above $T_{n}$. 
Free energy derivatives in temperature determined the transition  to be 
third-order. The bottleneck manifests as a ripple on the equilibrium free 
energy surface over nematic order and defect density. Temperature 
quench evolutions under EAMC dynamics  showed that the crossover 
was facilitated by {\it local} correlations of defect cores and nematic clusters. 
The system-wide distributions of director orientations calculated from 
EAMC simulations at different temperatures, exhibit macroscopic 
occupations of polar angle projections at $\pm 1/\sqrt{3}$ and 
azimuthal angle projections  at $\pm 1/\sqrt{2}$.

In analyzing this model, we find the micro-canonical inverse effective 
temperature and its derivatives form a common conceptual link between 
the inflexion diagnostic for finite-system transitions \cite{Gross,Schnabel,Qi}, 
the entropy barrier crossings of the partial equilibration scenario \cite{Ritort,Bonilla,Garriga,Crisanti04,Crisanti13,ShankarEPL}, and 
general entropy barrier acceptance probabilities \cite{WL,LW,Lbook}.

Usually, phase transitions have been associated either with order 
parameters from long range correlations, or with topological defect 
unbinding with power-law correlations. This model finds, that for 
narrow entropic bottlenecks constraining access to transitions, 
short-range correlations can emerge between order parameters and 
defects, to facilitate crossovers through a configurational sparsity
 gap between disordered and ordered phases. 

  Further work could apply the EAMC protocols, $\beta$-derivative analysis, 
  and quench simulations to  other problems, including frustrated 
  Heisenberg antiferromagnets \cite{Kawamura1,Kawamura2}, 
  biaxial liquid crystals \cite{BKL18, BKL20, BKL21, BKLThesis}, 
  interacting molecules \cite{Shell} of a glassy melt, and protein 
  folding controlled by  entropic bottlenecks of phase space 
  golf-holes \cite{ShankarEPL,Shakhnovich, Wolynes,Udgaonkar}.

\begin{acknowledgments}
We acknowledge computational support from the Centre for Modelling 
Simulation and Design (CMSD) at the University of Hyderabad. B.K.L.
acknowledges financial support from the Department of Science and 
Technology, Government of India vide Grant No. DST/WOS-A/PM-4/2020  
to carry out this work. We thank Mustansir Barma for useful conversations.
\end{acknowledgments}

 \end{document}